\documentclass[prx,superscriptaddress,twocolumn]{revtex4-2}
\usepackage[utf8]{inputenc}

\usepackage{verbatim}
\usepackage{graphicx}
\usepackage{amsmath}
\usepackage{amsfonts}
\usepackage{amsthm}
\usepackage{amssymb}
\usepackage{amsbsy}
\usepackage{wasysym}
\usepackage{bm}
\usepackage{mathrsfs}
\usepackage{color}
\usepackage{hyperref}
\usepackage{soul}
\usepackage{bbm}
\usepackage{physics}
\usepackage{orcidlink}

\hypersetup{
    colorlinks=true,
    linkcolor=red,        
    citecolor=blue,
    breaklinks=true,
    urlcolor=blue
}
\newcommand{\bo}{\mathbf{\Omega}}

\newcommand{\sn}{\mathrm{sn}}
\newcommand{\cn}{\mathrm{cn}}
\newcommand{\dn}{\mathrm{dn}}

\begin{document}

\title{Asymmetric decay of quantum many-body scars in XYZ quantum spin chains}

\author{Dhiman Bhowmick}
\affiliation{Department of Physics, National University of Singapore, Singapore 117551}

\author{Vir B.~Bulchandani}
\affiliation{Department of Physics, National University of Singapore, Singapore 117551}
\affiliation{Department of
Physics and Astronomy, Rice University, 6100 Main Street
Houston, TX 77005, USA}

\author{Wen Wei Ho}
\altaffiliation{\href{mailto:wenweiho@nus.edu.sg}{wenweiho@nus.edu.sg}}
\affiliation{Department of Physics, National University of Singapore, Singapore 117551}
\affiliation{Centre for Quantum Technologies, National University of Singapore, 3 Science Drive 2, Singapore 117543}

\date{\today}

\begin{abstract} 
Quantum many-body scars are atypical energy eigenstates of chaotic quantum many-body systems that prevent certain special non-equilibrium initial conditions from thermalizing. We point out that quantum many-body scars  exist for {\it any} nearest-neighbor spin-$S$ XYZ quantum spin chain, and arise in the form of an infinite family of  highly excited yet nonentangled product-state eigenstates, which define periodic textures in spin space. This set of scars, discovered originally by Granovskii and Zhedanov in 1985, encompasses both the experimentally relevant `spin helices' for  XXZ chains and more complicated helix-like states constructed from Jacobi elliptic functions for generic XYZ chains. An appealing feature of  Granovskii-Zhedanov scars is that they are well-defined in the semiclassical limit  $S \to \infty$, which allows for a systematic and analytical treatment of their dynamical instability to perturbations of the Hamiltonian. Using time-dependent spin-wave theory, we predict that upon perturbing along certain directions in Hamiltonian space, Granovskii-Zhedanov scars exhibit a dramatic asymmetry in their decay: depending on the sign of the perturbation, the decrease of their contrast is either slow and linear, or fast and exponential in time. This asymmetry can be traced to the absence (presence) of imaginarity in the spectrum of the Bogoliubov Hamiltonian governing quantum fluctuations about the scar, which corresponds to the absence (presence) of a non-zero Lyapunov exponent for the limiting  classical trajectory. Numerical simulations using matrix product states (MPS) and infinite time-evolving block decimation (iTEBD) confirm that our prediction remains valid even far from the semiclassical limit. Our findings challenge existing theories of how quantum-many body scars relax.
\end{abstract}

\maketitle


\section{Introduction}

Understanding thermalization and its breakdown in the dynamics of isolated quantum many-body systems is a foundational problem in modern physics, with conceptual and practical implications ranging from the validity of statistical mechanics to the control of large-scale quantum devices. By now, a number of physical mechanisms that prevent thermalization have been uncovered, such as quantum integrability~\cite{korepin1997quantum}, many-body localization~\cite{Basko_2006,OgHuse}, and Hilbert space fragmentation~\cite{Sala_2020,Khemani_2020}. In these examples, ergodicity is broken strongly: generic initial configurations will fail to thermalize. 

Quantum many-body scars (QMBS) represent a striking intermediate scenario wherein ergodicity is only weakly broken, analogous to mixed phase spaces in classical mechanics~\cite{gutzwiller2013chaos}. First identified in experiments on Rydberg quantum simulators~\cite{bernien2017probing,turner2018weak}, quantum many-body scars are atypical, non-ergodic  eigenstates of otherwise chaotic quantum many-body Hamiltonians that generate non-thermalizing dynamics when the system is initialized in certain special non-equilibrium initial states. 
They were named 
by analogy with quantum scars in single-particle systems  --- exceptional wavefunctions that are not uniformly distributed in phase space but instead concentrate on weakly unstable periodic classical orbits~\cite{heller1984bound} --- and indeed a number of works have since
attempted to connect quantum many-body scars to dynamics in a suitably defined classical many-body phase space~\cite{Ho_2019,turner2021correspondence,Cotler_2023,hummel2023genuine,lerose2023theory,ermakov2024periodic,evrard2024quantum,pizzi2024quantumscarsmanybodysystems,muller2024semiclassicaloriginsuppressedquantum,omiya2024quantummanybodyscarsremnants}. The terminology of quantum many-body scars has since expanded to  include a vast array of non-thermal eigenstates with largely non-classical origins, ranging from projector embeddings~\cite{Shiraishi_2017, PhysRevLett.122.220603}, to group theory~\cite{pakrouski2020many,O_Dea_2020}, to commutant algebras~\cite{moudgalya2023exhaustivecharacterizationquantummanybody}.

While the construction and classification of quantum many-body scars has been studied in some detail, their dynamical stability is less well understood. Specifically, it is expected that perturbations of a Hamiltonian hosting quantum many-body scarring will cause the system to relax and thermalize even when it is prepared in the originally non-thermalizing configuration associated with scars, due to the latter's immediate hybridization with neighboring thermal energy eigenstates. However, a systematic theory yielding the timescales and temporal profiles of the ensuing  thermalization dynamics is lacking. Although several works illustrate slow decay in specific perturbed scarred models~\cite{SlowThermalization_SPXP, Fidelity_Succeptibility_SPXP, DistortedScar_SPXP, DisorderDetection_QSimulator_SPXP, TwoDimensionalPXP_SPXP,Stable_Fermionic_Scars,Gauge_Theory}, most of these rely on a perturbative, short-time expansion of the relaxation dynamics; an exception is Ref.~\cite{SlowThermalization_SPXP} which provides general (but potentially loose) constraints on thermalization timescales, using bounds on quantum information propagation.

\begin{figure*}[t]
\centering
\includegraphics[width=0.99\textwidth]{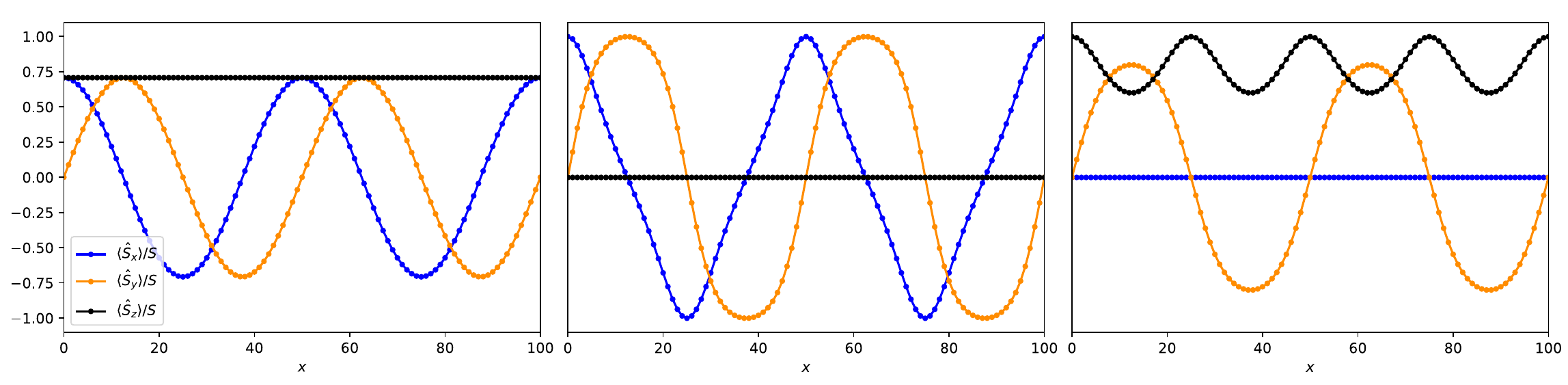}
\caption{Illustrative examples of spin textures for the various kinds of Granovskii-Zhedanov scars considered in this paper. In all cases we set $L=100$, the wavenumber $q = 8K(\kappa)/L$, and $\phi = 0$. \textbf{Left}: an ordinary transverse spin helix defined by $\kappa=0$ and $\gamma=\sqrt{2}/2$, corresponding to an exact product state eigenstate in the spin-$S$ XXZ chain with $J_x=J_y=1, \, J_z=0.992...$, \textbf{Middle}: a generalized transverse spin helix winding in the $x$-$y$ plane defined by $\kappa=0.9$ and $\gamma=0$, corresponding to an exact product state eigenstate in the spin-$S$ XYZ chain with $J_x=0.986..., \,J_y=1, \,J_z= 0.983...$, \textbf{Right}: a generalized longitudinal spin helix winding in the $y$-$z$ plane defined by $\kappa=0.8$ and $\gamma=1$, corresponding to an exact product state eigenstate in the spin-$S$ XYZ chain with $J_x=0.991..., J_y=1, J_z=0.987...$. }
\label{fig:GZprofiles}
\end{figure*}

In this paper, we identify a family of Hamiltonians with associated quantum many-body scars for which a theory of their dynamical stability at late times can be systematically and analytically developed, and which goes beyond previous studies. 
Concretely, we first point out that {\it any} nearest-neighbor, spin-$S$, XYZ quantum spin chain  admits an infinite family of highly-excited yet non-entangled  product-state eigenstates, which define periodic textures in spin space (see Fig.~\ref{fig:GZprofiles}). 
For example, in the special case of XXZ chains, these are ``transverse spin-helix states'': spin textures winding tranversely along the chain whose wavevectors have a definite relation to the model anisotropy, which have recently been theoretically studied~\cite{popkov2021phantom} and experimentally probed with ultracold  atoms~\cite{jepsen2021transverse}~(note generalizations exist for higher spatial dimensions~\cite{Jepsen_2022}). For general XYZ chains, the corresponding product-state eigenstates describe spin textures that are generically also helix-like but have more complicated structures, governed mathematically by Jacobi elliptic functions. Remarkably, this perhaps surprising result --- that {\it all} XYZ quantum spin chains harbor exact product state scars --- was already discovered by Granovskii and Zhedanov back in 1985~\cite{GZEarlier,granovskii1985periodic}, decades before the field of quantum many-body scarring was initiated~\cite{turner2018weak} or the works \cite{popkov2021phantom, jepsen2021transverse, Jepsen_2022}, and we will henceforth refer to them as  Granovskii-Zhedanov (GZ) scars.

We focus on the quench dynamics of GZ scars following a sudden perturbation of the underlying  parent Hamiltonian and study how they relax over time. 
A virtue of GZ scars is that they are well-defined in the semiclassical  limit $S \to \infty$, and further have classical mean-field dynamics that is analytically understood for perturbations along certain directions in Hamiltonian space. 
This allows us to cleanly separate their full quantum dynamics into the classical mean-field dynamics and quantum fluctuations above this, the latter of which can be captured by a controlled $1/S$ expansion in time-dependent spin-wave theory (i.e., a semiclassical expansion). 
We predict a dramatic asymmetry in how the GZ scars of generic XYZ quantum spin chains respond to perturbations along certain directions in Hamiltonian space: depending on the {\it sign} of the perturbation, their decay is either slow and linear, or fast and exponential.  A comparison of our predictions to tensor-network-based numerical simulations reveals that they remain valid even far from the semiclassical limit.
We identify a specific criterion 
for such behavior to arise: we find that asymmetric decay in the {quantum} dynamics in the semiclassical {\it regime} is captured by imaginarity in the spectrum of the spin-wave Hamiltonian governing fluctuations about the quantum scar in a $1/S$ expansion; 
further, we show this coincides with the onset of modulational instability~\cite{benjamin1967disintegration,kivshar1993peierls,lakshmanan2008dynamic} of the periodic orbit in the  semiclassical {\it limit}, captured by the absence or presence of a non-zero Lyapunov exponent in the Floquet spectrum of the classical Hamiltonian governing perturbations about the orbit.

Our work advances the conceptual understanding of how quantum many-body scars may relax, unveiling a class of scars whose relaxation dynamics is \textit{asymmetric} and {\it non-analytic} in the perturbation strength, 
behavior which lies beyond existing short-time or perturbative techniques.
Furthermore, we expect the detailed stability theory we have developed for the GZ scars in spin-$S$ XYZ quantum spins, which establishes a quantitative quantum-classical connection between semiclassical quantum fluctuations and modulational instability of classical orbits, to hold more generally for other quantum many-body systems hosting quantum many-body scars that are well-defined in the semiclassical limit.
%
This importantly brings the theory of quantum many-body scars closer in spirit to the established theory of single-particle quantum scars by  E.~Heller~\cite{heller1984bound}.

\section{Granovskii-Zhedanov scars in spin-$S$ XYZ quantum spin chains}

In this work, we consider     nearest-neighbor, spin-$S$, XYZ quantum spin chains in 1D: 
\begin{equation}
\label{eq:XYZintro}
\hat{H}_{XYZ} = \sum_{j=1}^L J_x \hat{S}_j^x \hat{S}_{j+1}^x + J_y \hat{S}_j^y \hat{S}_{j+1}^y + J_z \hat{S}_j^z \hat{S}_{j+1}^z.
\end{equation}
Above, $\hat{S}_j^x, \hat{S}_j^y, \hat{S}_j^z$ are   spin-$S$ operators and $J_x, \, J_y, \, J_z$ are a set of real numbers parameterizing the strengths of interaction along different directions; it is expected that these Hamiltonians are non-integrable in general~\footnote{That is, staying away from the $S=1/2$ case, where it is known to be integrable.}. 
We note that an XYZ quantum Hamiltonian may be experimentally realized through Floquet driving of XXZ-type Hamiltonians~\cite{PhysRevX.10.031002, PhysRevX.14.031017}, which are naturally realized in cold atom and nitrogen-vacancy quantum simulators, for example.

We first claim that  {\it any} XYZ Hamiltonian is, in fact,   quantum many-body scarred: precisely,  in the thermodynamic limit $L \to \infty$, for {\it any} spin $S$ and for {\it any} values of the parameters $J_x, \, J_y, \, J_z$, the Hamiltonian Eq.~\eqref{eq:XYZintro} admits an infinite family of exact and high-energy yet completely unentangled product-state eigenstates, which are hence also the special initial conditions from which the system would fail to thermalize.

To write down these quantum many-body scars explicitly, let us without loss of generality set $J_y = 1$ and take $0 \leq J_z \leq J_x \leq 1$ \footnote{Note we can restrict the parameters to this range because firstly the sign of the Hamiltonian is irrelevant for the purposes of establishing an energy eigenstate, and secondly we may use a rotation on every other site to transform the Hamiltonian into one where every parameter $J_x, J_y, J_z$ is of the same sign.}. Define
\begin{align}
\kappa^2= \frac{1-J_x^2}{1-J_z^2},
\label{eqn:kappa_squared}
\end{align}
which takes values in $[0,1]$, and find (the non-unique) $q \in \mathbb{R}$ for which
\begin{align}
\cn(q,\kappa) = J_z, \dn(q,\kappa) = J_x.
\label{eqn:q_eqn}
\end{align} 
Here, $\sn(q,\kappa), \cn(q,\kappa), \dn(q,\kappa)$ are the Jacobi elliptic functions \cite{weisstein2002jacobi}, with whose elliptic modulus $\kappa$ has been identified.
Then, the statement is that the product of Bloch coherent states 
\begin{align}
|\kappa,q,\gamma,\phi\rangle:=
\otimes_{j=1}^L | \mathbf{\Omega}_j\rangle,
\label{eqn:scar_states}
\end{align}
for {\it any}   $\gamma \in [0,1]$ and {\it any} $\phi \in \mathbb{R}$, 
where each local state $|\mathbf{\Omega}_j\rangle$ is  defined by the normalized Bloch vector 
\begin{align}
\mathbf{\Omega}_j & =(\langle\mathbf{\Omega}_j|\hat{S}_j^x|\mathbf{\Omega}_j\rangle/S, \langle\mathbf{\Omega}_j|\hat{S}_j^y|\mathbf{\Omega}_j\rangle/S, \langle\mathbf{\Omega}_j|\hat{S}_j^z|\mathbf{\Omega}_j\rangle/S)^T  \nonumber \\
& = 
\begin{pmatrix} \alpha\, \mathrm{cn}(qj+\phi,\kappa) \\
\beta\, \mathrm{sn}(qj+\phi,\kappa) \\
\gamma\, \mathrm{dn}(qj+\phi,\kappa)\end{pmatrix}
\label{eqn:scar_state}
\end{align}
with $\alpha = \sqrt{1-\gamma^2}$ and $\beta = \sqrt{1-\gamma^2(1-\kappa^2)}$, is an exact eigenstate of the XYZ Hamiltonian Eq.~\eqref{eq:XYZintro} for any spin $S$. It has energy density 
\begin{align}
\epsilon&=\lim_{L\to\infty}\frac{E}{L}
\nonumber \\
&= S^2\left(\cn(q,\kappa) \dn(q,\kappa) +  \sn(q,\kappa) \left[\varepsilon(q,\kappa) - \frac{ q \varepsilon(K(\kappa),\kappa)}{K(\kappa)} \right]\right)
\end{align}
in the thermodynamic limit,  where $\varepsilon(u,\kappa) = \int_0^u dv \, \dn^2(v,\kappa)$ denotes Jacobi's epsilon function and $K(\kappa)$ is the complete elliptic integral of the first kind~\cite{granovskii1985periodic, weisstein2002jacobi} (note that $\epsilon$ is independent of $\gamma$ and $\phi$).

We may visualize these product-state eigenstates by plotting the behavior of the Bloch vectors $\bo_j$; in general (for $\kappa \neq 1$), they describe   intricate periodic structures in spin space moving along the chain, see Fig.~\ref{fig:GZprofiles}~\footnote{Note the Jacobi elliptic functions $\{\cn(x,\kappa), \sn(x,\kappa),\dn(x,\kappa)\}$ are periodic in $x$ with a joint wavelength of $4K(\kappa)$, where $K(\kappa)$ is the complete elliptic integral of the first kind~\cite{weisstein2002jacobi}.}.
A simple and illustrative example occurs when $J_x = 1$ (XXZ anisotropy) resulting in $\kappa = 0$ 
and hence the Jacobi elliptic functions $\{\cn,\sn,\dn\}$ reducing to basic trigonometric functions $\{\cos,\sin,1\}$; these describe so-called `transverse spin-helix scar states' of XXZ chains with a Bloch spin texture spiraling in the $x$-$y$ transverse plane with wavenumber $q = \cos^{-1}(J_z)$ and  longitudinal component governed by $\gamma$ (Fig.~\ref{fig:GZprofiles}, left),
which have 
  recently been studied theoretically  
\cite{popkov2021phantom} and experimentally in ultracold  atoms~\cite{jepsen2021transverse, Jepsen_2022}.
For general values of system parameters such that $\kappa \neq 1$,
Eq.~\eqref{eqn:scar_state} still describes  helical patterns with wavenumber $q$ and wavelength $4K(\kappa)/q$, but with more complicated-looking profiles  (see Fig.~\ref{fig:GZprofiles}, middle and right).
Note when the system size $L$ is finite and assuming the chain is closed into a ring with periodic boundary conditions, an additional commensurability condition on the wavenumber $q$  has to be satisfied:
\begin{align}
q = 4MK(\kappa)/L
\end{align}
for some integer $M$. 

The claim that there is an entire family of product states of the form Eq.~\eqref{eqn:scar_states},  which are exact high-energy eigenstates of a given XYZ Hamiltonian, constitutes a perhaps surprising statement of the {\it ubiquity} of quantum many-body scars in quantum spin chains, a fact that seems not to have been appreciated in the literature thus far.
Remarkably, these exact product state eigenstates were discovered already in 1985 by Granovskii and Zhedanov \cite{GZEarlier,granovskii1985periodic} --- not in the context of quantum many-body scarring as their work preceded the establishment of the field by decades~\cite{turner2018weak} but rather from the point of view of integrable structures in spin chains --- and we will henceforth refer to them  as Granovskii-Zhedanov (GZ) scars. We provide an outline of the proof of this claim in Appendix \ref{appendix:proof}.

Because GZ scars exist for any spin-$S$, and in particular in the semiclassical limit $S \to \infty$, we may study their quantum dynamics using well-controlled semiclassical methods, which provides a unique opportunity to develop a systematic theory of their late-time dynamical stability to perturbations. 
We will develop this theory in Sec.~\ref{sec:gentheory}.
This is in contrast to other quantum many-body scars established in the literature as arising from Hamiltonian constructions using projector embeddings, group theory or commutant algebras~\cite{Shiraishi_2017, PhysRevLett.122.220603, pakrouski2020many,O_Dea_2020, moudgalya2023exhaustivecharacterizationquantummanybody}, which are typically defined far from any notion of semiclassical limit, 
and are therefore less amenable to a comprehensive, analytical understanding of their dynamics.

\subsection{Dynamical stability of GZ scars}
The main physical question we address in this work is the dynamical stability of GZ scars following a sudden perturbation of their underlying parent quantum many-body Hamiltonian
\begin{align}
\hat{H}_{XYZ} \mapsto \hat{H} = \hat{H}_{XYZ} + \delta \hat{H},
\label{eqn:perturbed_H}
\end{align}
i.e.,~quench dynamics   $|\psi(t)\rangle = e^{-i \hat{H} t} |\psi(0)\rangle$ where the initial state $|\psi(0)\rangle$ is the associated GZ scar $|\kappa,q,\gamma,\phi\rangle$  of $\hat{H}_{XYZ}$ . For simplicity, we henceforth set the phase of the GZ scar $\phi = 0$ and refer to it as $|\kappa,q,\gamma\rangle$.
Further, we will consider perturbations $\delta \hat{H}$ of the form
\begin{align}
\delta \hat{H} = \sum_{j=1}^L \delta J_x \hat{S}_j^x \hat{S}_{j+1}^x + \delta J_y \hat{S}_j^y \hat{S}_{j+1}^y + \delta J_z \hat{S}_j^z \hat{S}_{j+1}^z;
\end{align}
that is, the perturbed Hamiltonian $\hat{H}$ is also an XYZ Hamiltonian but with  interaction strengths changed $(J_x,J_y,J_z)\mapsto (J_x,J_y,J_z)+(\delta J_x, \delta J_y, \delta J_z)$,
such that the  GZ state $|\kappa,q,\gamma\rangle$ is generically no longer an eigenstate and undergoes nontrivial dynamics.

To diagnose the stability of a quantum state in dynamics, a quantity  often studied is the fidelity or return probability $\mathcal{F}(t) = |\langle \psi(t)| \psi(0)\rangle|^2$. However, in the present case, because the GZ scars are mean-field states
--- being unentangled products of Bloch states ---
we can choose to study instead their stability about their dynamics in the semiclassical limit  $S \to \infty$. Precisely, what we mean is this: in the semiclassical limit, the dynamics of each spin $j$, 
whose state space is fully captured by a normalized Bloch coherent vector $\bo_j(t)$, 
will be governed by the discrete Landau-Liftshitz mean-field equations of motion,
\begin{equation}
\label{eq:classEoM}
\dot{\bo}_j(t) = \frac{\partial H^{\mathrm{cl}}}{\partial \bo_j}(t) \times \bo_j(t)
\end{equation}
where $H^{\mathrm{cl}}$ is the classical Hamiltonian associated with the perturbed quantum Hamiltonian $\hat{H}$. 
We   then define the `contrast'
\begin{align}
\mathcal{D}(t) &= \frac{1}{LS}\sum_{j=1}^L \langle \psi(t) | \bo_j(t) \cdot \hat{\mathbf{S}}_j | \psi(t)\rangle,
\label{eq:defcontrast_0}
\end{align}
which amounts to measuring the component of the time-evolved quantum state $|\psi(t)\rangle$ along the direction of the mean-field trajectory ${\bo}_j(t)$. We note this is a much more experimentally accessible quantity than the fidelity. 
By construction, $\mathcal{D}(t) = 1$ in the   limit $S \to \infty$, but for finite $S$, the contrast will be expected to decay.
The contrast Eq.~\eqref{eq:defcontrast_0}    generalizes the `spin contrast' which has been studied recently in ultracold atom experiments probing the relaxation of transverse spin-helix states in XXZ chains~\cite{Jepsen_2020,jepsen2021transverse,Jepsen_2022}.

How should we expect the contrast $\mathcal{D}(t)$ to decay? 
A first guess would be to treat $\delta \hat{H}$ as a generic small perturbation. Then, one may argue that the population on the scar eigenstate $|\kappa,q,\gamma\rangle$ of the unperturbed Hamiltonian $\hat{H}_{XYZ}$, which initially starts at unity, gets depleted via coupling to neighboring eigenstates $|n\rangle$ of equal energy, mediated by $\delta \hat{H}$. This should occur at a rate $\Gamma$ captured by Fermi's golden rule, which involves matrix elements  
$|\langle n|\delta \hat{H}|\kappa,q,\gamma\rangle|^2$; thus, one may na\"ively   predict that the contrast $\mathcal{D}(t)$ decays at a rate $\Gamma \sim \|\delta \hat{H}\|^2$. This prediction exhibits two salient features: first, regardless of the kind of perturbation, the decay rate is {\it analytic} in the perturbation strength. 
Second, also regardless of the kind of perturbation,  
 the relaxation dynamics is {\it insensitive} to the {\it sign} of the perturbation strength $\delta \hat{H}$. 

As we shall argue in what follows, and which constitutes one of the key findings of our work, this seemingly reasonable analysis turns out to be inaccurate: depending on the type of GZ scar considered, there may exist perturbations along certain directions in Hamiltonian space where the contrast $\mathcal{D}(t)$ exhibits {\it non-analytic} behavior in the perturbation strength, and furthermore is {\it asymmetric} --- demonstrating a dramatic and surprising difference in relaxation dynamics depending on the sign of the perturbation. 
This  is captured by our general theory of quantum fluctuations about generic mean-field classical dynamics, developed in Sec.~\ref{sec:gentheory}. In Sec.~\ref{sec:scars_perturbation} we  identify three families of GZ scars and the perturbations of their parent Hamiltonians to which our theory can readily be applied and yields analytical results; Sec.~\ref{sec:TSH} and Sec.~\ref{sec:GSH} present the concrete predictions of our theory applied to these states,
which are further corroborated by numerics.

\section{Theory of quantum fluctuations about mean-field classical dynamics}
\label{sec:gentheory}

In this section, we develop a general theory to understand the relaxation dynamics of the contrast $\mathcal{D}(t)$ that holds not only for the GZ scars of interest, but in fact also  for {\it arbitrary} mean-field quantum states --- that is, product states of Bloch coherent states.  
Concretely, we will perform a linear stability analysis of mean-field states
using time-dependent spin-wave theory~\cite{ruckriegel2012time}, which captures their quantum dynamics due to quantum fluctuations on top of their classical trajectories, to leading order in $1/S$.
%

To set up our theory, consider the following 
general nearest-neighbor-interacting quantum spin-$S$ chain of the form
\begin{equation}
\label{eq:HXYZ}
\hat{H} = \sum_{j=1}^L \hat{\mathbf{S}}_j \cdot J \hat{\mathbf{S}}_{j+1},
\end{equation}
where $J$ is an arbitrary real, symmetric three-by-three exchange matrix. 
Additionally, periodic boundary conditions $\hat{\mathbf{S}}_{L+1} \equiv \hat{\mathbf{S}}_{1}$ are imposed.  
Note that the Hamiltonian Eq.~\eqref{eq:HXYZ} can always be expressed in the form Eq.~\eqref{eq:XYZintro} by rotating the spin operators to align with the principal axes of $J$ and rescaling, so encompasses our setting of interest. One may of course generalize the  present analysis to other classes of Hamiltonians e.g., those with longer range interactions or with spatial inhomogeneity.

Now, given a solution $\bo_j(t)$ to the mean-field $(S \to \infty)$ Landau-Lifshitz dynamics  associated with Eq.~\eqref{eq:HXYZ}, which we term the `semiclassical trajectory' (see Appendix \ref{app:quantclass}), which satisfies
\begin{equation}
\label{eq:MFEoM}
\dot{\mathbf{\Omega}}_j(t) = SJ(\mathbf{\Omega}_{j-1}(t)+\mathbf{\Omega}_{j+1}(t))\times \mathbf{\Omega}_j(t),
\end{equation}
we would like to understand how the product of the corresponding Bloch coherent states prepared at time $t = 0$,
\begin{align}
|\vec{\bo}(0)\rangle = \otimes_{j=1}^L | \mathbf{\Omega}_j(0)\rangle,
\end{align}
evolves under
 the full quantum dynamics induced by $\hat{H}$.
 We consider the contrast
 \begin{align}
\mathcal{D}(t) &= \frac{1}{LS}\sum_{j=1}^L \langle \psi(t) | \bo_j(t) \cdot \hat{\mathbf{S}}_j | \psi(t)\rangle
\end{align}
as the relevant object of interest, which as mentioned before, measures the   stability of the mean-field dynamics to quantum fluctuations away from the $S \to \infty$ limit. 
 

\subsection{The rotating frame}
We first decompose the exact Schr{\"o}dinger-evolved state vector $|\psi(t)\rangle = e^{-i\hat{H}t} |\vec{\bo}(0)\rangle$ into (i) the purely classical mean-field dynamics 
$ |\vec{\bo}(t)\rangle$, 
where $|\vec{\bo}(t)\rangle$ is the time-evolved quantum state comprising an unentangled product of Bloch states with Bloch vectors evolving via Eq.~\eqref{eq:MFEoM}, and (ii) residual quantum dynamics above it. 


To this end, we employ
a gauge transformation into the rotating frame of reference defined by the classical dynamics. Specifically, we define three-by-three rotation matrices $R_j(t)$ with the property that $\bo_j(t) = R_j(t) \hat{\mathbf{z}}$, where $\hat{\mathbf{z}}$ denotes the unit vector along the $z$-axis. Similarly we define their spin-$S$ representations $\hat{\mathcal{R}}_j(t)$ (quantum operators) such that $\hat{\mathcal{R}}_j(t) (\mathbf{a}_j \cdot \hat{\mathbf{S}}_j) \hat{\mathcal{R}}^\dagger_j(t) = (R_j(t)\mathbf{a}_j) \cdot \hat{\mathbf{S}}_j$ for all $\mathbf{a}_j\in\mathbb{R}^3$. This yields a quantum rotation operator $\hat{\mathcal{R}}(t) = \prod_{j=1}^L \hat{\mathcal{R}}_j(t)$ that maps the reference fully $z$-polarized ferromagnetic state $\ket{\Uparrow} = \otimes_{j=1}^L |\hat{\mathbf{z}}\rangle$ to the instantaneous mean-field spin texture $|\vec{\bo}(t)\rangle$, i.e. $
|\vec{\bo}(t)\rangle = \hat{\mathcal{R}}(t)\ket{\Uparrow}$.

In terms of this operator, the Schr{\"o}dinger-picture state vector can then be written as
\begin{align}
|\psi(t)\rangle = \hat{\mathcal{R}}(t)\hat{U}_{\mathrm{R}}(t) \ket{\Uparrow},
\label{eqn:decomp}
\end{align}
 where the rotated time-evolution operator $\hat{U}_{\mathrm{R}}(t) = \hat{\mathcal{R}}^\dagger(t) e^{-i\hat{H}t} \hat{\mathcal{R}}(0)$ captures the residual quantum fluctuations about the mean-field solution. The latter is generated by an effective, time-dependent `rotating Hamiltonian'
\begin{align}
\nonumber
\hat{H}_{\mathrm{R}}(t) &= i\frac{d}{dt}\hat{U}_{\mathrm{R}}(t) \hat{U}^\dagger_{\mathrm{R}}(t)\\
&= \hat{\mathcal{R}}^\dagger(t) \hat{H}\hat{\mathcal{R}}(t) - i \hat{\mathcal{R}}^\dagger(t) \frac{d}{dt}\hat{\mathcal{R}}(t),
\end{align}
so that $\hat{U}_R(t)$ is expressible as $\hat{U}_R(t) = \mathcal{T}e^{-i\int_0^t \hat{H}_\text{R}(t') dt'}$. 

 For $\hat{H}$ as in Eq.~\eqref{eq:HXYZ}, the rotating Hamiltonian $\hat{H}_{\mathrm{R}}(t)$ reads
\begin{equation}
\hat{H}_{\mathrm{R}}(t) = \sum_{j=1}^L \hat{\mathbf{S}}_{j}^T J_{\mathrm{R},j}(t) \hat{{\mathbf{S}}}_{j+1} + \mathbf{h}_{\mathrm{R},j}(t) \cdot \hat{\mathbf{S}}_j,
\end{equation}
where the transformation to the rotating frame has induced effective exchange matrices
\begin{equation}
J_{\mathrm{R},j}(t) := R^T_j(t)J R_{j+1}(t)
\end{equation}
and an effective magnetic field that satisfies
\begin{equation}
\mathbf{h}_{\mathrm{R},j}(t) \times \mathbf{v}= -R^T_j(t) \frac{d}{dt}R_j(t) \mathbf{v}
\end{equation}
for all $\mathbf{v} \in \mathbb{R}^3$, both of which are generically space- and time-dependent.

Note the condition that $\mathbf{\Omega}_j(t)$ satisfies the Landau-Lifshitz equations Eq.~\eqref{eq:MFEoM} imposes the non-trivial constraint
 \begin{equation}
\label{eq:statcond}
S(J_{\mathrm{R},j-1}^T(t) + J_{\mathrm{R},j}(t))\hat{\mathbf{z}} + \mathbf{h}_{\mathrm{R},j}(t) = \omega_j(t) \hat{\mathbf{z}}
\end{equation}
of stationarity in the rotating frame, where we defined the constant of proportionality
\begin{equation}
\label{eq:defomega}
\omega_j(t) = S(J_{\mathrm{R},j-1}^{zz}(t) + J_{\mathrm{R},j}^{zz}(t)) + h^z_{\mathrm{R},j}(t),
\end{equation}
see Appendix \ref{app:quantclass} for details.

\subsection{Time-dependent spin-wave Hamiltonian}
According to time-dependent spin-wave theory~\cite{ruckriegel2012time}, the quantum fluctuations about the mean-field trajectory $|\vec{\bo}(t)\rangle$, i.e., the dynamics generated by $\hat{H}_\text{R}(t)$, can   be obtained order-by-order in $1/S$ by applying conventional spin-wave theory.

We invoke the
Holstein-Primakoff transformation from spins to bosonic operators, which reads $\hat{S}^+_j = \sqrt{2S}\hat{a}_j, \, \hat{S}^-_j = \sqrt{2S}\hat{a}_j^\dagger, \, \hat{S}^z_j =  S-\hat{a}_j^\dagger \hat{a}_j$  at order $S^0$. By the constraint Eq.~\eqref{eq:statcond},  the secular terms are seen to vanish, yielding
\begin{equation}
\label{eq:SWapprox}
\hat{H}_{\mathrm{R}}(t) = \sum_{j=1}^L \left(S^2 J_{\mathrm{R},j}^{zz}(t)+Sh^z_{\mathrm{R},j}(t)\right) + \hat{H}_{\mathrm{SW}}(t) + \mathcal{O}(S^{1/2}),    
\end{equation}
where the first term is simply a time-dependent energy shift and the second term is the effective spin-wave Hamiltonian
\begin{align}
\nonumber
&\hat{H}_{\mathrm{SW}}(t) = \sum_{j=1}^L \eta_j(t) \hat{a}_{j+1}^\dagger \hat{a}_j + \eta_j^*(t) \hat{a}_j^\dagger \hat{a}_{j+1} \\
\label{eq:HSW}
&+\zeta_j(t) \hat{a}_{j+1}^\dagger \hat{a}_j^\dagger + \zeta_j^*(t) \hat{a}_{j} \hat{a}_{j+1} + V_j(t) \hat{a}^\dagger_j \hat{a}_j,
\end{align}
a quadratic bosonic Hamiltonian with   local hopping strength, pair-creation amplitude and onsite potential given by
\begin{align}
\label{eq:SWcoupl1}
\eta_j &= \frac{S}{2}\left(J_{\mathrm{R},j}^{xx}(t) + J_{\mathrm{R},j}^{yy}(t) + i(J^{xy}_{\mathrm{R},j}(t)-J^{yx}_{\mathrm{R},j}(t))\right),\\
\label{eq:SWcoupl2}
\zeta_j &= \frac{S}{2}\left(J_{\mathrm{R},j}^{xx}(t) - J_{\mathrm{R},j}^{yy}(t) + i(J^{xy}_{\mathrm{R},j}(t)+J^{yx}_{\mathrm{R},j}(t))\right),\\
\label{eq:SWcoupl3}
V_j &= -S(J^{zz}_{R,j-1}(t) + J^{zz}_{R,j}(t)) - h_{\mathrm{R},j}^z(t). 
\end{align}
The dynamics generated by the truncation of $\hat{H}_R(t)$ to terms up to $\hat{H}_\text{SW}(t)$ is referred to as the `spin-wave approximation'. 

We observe from Eq.~\eqref{eqn:decomp} that because the initial state in the rotating frame is the fully $z$-polarized ferromagnetic state  $\ket{\Uparrow} = \otimes_{j=1}^L |\hat{\mathbf{z}}\rangle$, at least as far as approximating the dynamics of $\hat{H}_R(t)$ within the spin-wave approximation goes, 
there is nontrivial quantum dynamics induced if and only if the pair-creation/annihilation amplitudes of $\hat{H}_{\mathrm{SW}}(t)$ do not vanish: $\zeta_j \neq 0$. 
In other words, the dynamical stability (or instability) of the mean-field state to quantum fluctuations, to linear order, is  governed crucially by the nature of the spin-wave Hamiltonian $\hat{H}_{\mathrm{SW}}(t)$.

\subsection{Quantum fluctuations about the mean-field trajectory probe classical modulational instabilities} 

We interrupt our analysis of the contrast here with an important conceptual point.
Thus far, our analysis has amounted to analyzing the quantum fluctuations {\it on top} of the mean-field trajectory $\bo_j(t)$, captured within the spin-wave approximation by the spin-wave Hamiltonian $\hat{H}_\text{SW}(t)$. This is the dynamical behavior of a mean-field state in the {\it semiclassical regime} (large but finite $S$). Intriguingly, we point out that very same physics 
is encoded in the Landau-Lifshitz equations of motion valid in the {\it semiclassical limit} ($S \to \infty$), namely in the modulational instability {\it of the} mean-field trajectory.
 Although this correspondence may appear obvious to some readers (and has in fact been employed in a previous work~\cite{Rodriguez_Nieva_2022}),  the statement in its most general form does not appear to have been reported cleanly in a concrete fashion; we do so here. 


Let $\bo_j(t)$ be an exact mean-field solution  and consider perturbations of the initial conditions:
\begin{align}
\bo_j(0) \mapsto \bo_j(0)+\delta \bo_j(0).
\end{align}
We then solve for the dynamics generated by the Landau-Liftshitz equations with this new initial condition. 
In the rotating frame where   the unperturbed Bloch vector always points in the $\hat{\mathbf{z}}$ direction by construction, we may decompose the new Bloch vector $\mathbf{v}_j(t)$ of each spin into the unperturbed part $\hat{\mathbf{z}}$ and the perturbation $\mathbf{w}_j(t)$, 
%
\begin{align}
\mathbf{v}_j(t) = \hat{\mathbf{z}} + \mathbf{w}_j(t),
\end{align}
where $R_j(t) \mathbf{w}_j(t) = \delta \bo_j(t)$ and $\hat{\mathbf{z}}\cdot \mathbf{w}_j(t) = 0$ by the normalization constraint. We can thus further express the latter as
\begin{align}
\mathbf{w}_j(t) = \begin{pmatrix}
q_j(t) \\ p_j(t) \\0
\end{pmatrix}.
\end{align}
Now we can ask: what are the equations of motion that $(q_j(t),p_j(t))$ satisfy?
Defining the complex variables $\alpha_j = \frac{1}{\sqrt{2}}(q_j + i p_j)$, we show  in Appendix \ref{app:quantclass} that upon linearization of the equations of motion, $\alpha_j(t)$  are solutions of Hamilton's equations of motion 
\begin{equation}
i\dot{\alpha}_j = \frac{\partial H^{\mathrm{cl}}_{\mathrm{SW}}}{\partial \alpha^*_j},
\end{equation}
where 
\begin{align}
\label{eq:classSW_main}
& H^{\mathrm{cl}}_{\mathrm{SW}}(t) =  \sum_{j=1}^L \eta_j(t) \alpha_{j+1}^*\alpha_j + \eta_j^*(t) \alpha_j^* \alpha_{j+1} \nonumber \\
 & + \zeta_j(t) \alpha^*_j \alpha^*_{j+1} + \zeta_j^*(t) \alpha_j \alpha_{j+1} + V_j(t)\alpha_j^* \alpha_j
\end{align}
is a classical Hamiltonian with coefficients $\{\eta_j(t),\zeta_j(t),V_j(t)\}$   exactly as in Eqs.~\eqref{eq:SWcoupl1}-\eqref{eq:SWcoupl3}.  We see that Eq.~\eqref{eq:classSW_main} is nothing more than the classical analog of the quantum spin-wave Hamiltonian Eq.~\eqref{eq:HSW} upon identifying bosonic operators $\hat{a}_j$ with phase space variables $\alpha_j$.
This is our claimed ``quantum-classical correspondence'': the spectrum of the quantum Hamiltonian $\hat{H}_\text{SW}$ governing fluctuations coincides with the spectrum of the classical Hamiltonian  $\hat{H}_\text{SW}^\text{cl}$ governing instability of classical trajectories. 

In particular, it follows that whenever the underlying mean-field solution $\bo_j(t)$ is periodic in time, the   Floquet spectrum of the spin-wave Hamiltonian Eq.~\eqref{eq:HSW}  governing the quantum fluctuations on top of the classical periodic orbit $\bo_j(t)$ will coincide with  the
 Floquet-Lyapunov spectrum governing the modulational instability of the classical periodic orbit $\bo_j(t)$. 
 Though we have derived it only for the class of models Eq.~\eqref{eq:HXYZ}  in this work, we expect this relation to hold more generally. 
Such a quantum-classical correspondence is important for connecting mean-field dynamics to Heller's conception of quantum scarring in terms of unstable periodic orbits~\cite{heller1984bound}, and we elaborate on this in the discussion section, Sec.~\ref{sec:discussion}. 

\subsection{Dynamics of the contrast in the linear stability regime}
\label{subsec:contrastdyn}

Returning now  to the contrast of the quantum state evolved from a given mean-field configuration  $|\vec{\bo}(0)\rangle$,  which can be exactly represented in the rotating frame as 
\begin{align}
\mathcal{D}(t) =
\frac{1}{LS} \sum_{j=1}^L \bra{\Uparrow} \hat{U}_{\mathrm{R}}^\dagger(t) \hat{S}_j^z \hat{U}_{\mathrm{R}}(t) \ket{\Uparrow},
\label{eq:defcontrast}
\end{align}
%
%
we can make the spin-wave theory approximation  (neglecting an unimportant global phase from Eq.~\eqref{eq:SWapprox})
\begin{align}
\hat{U}_\mathrm{R}(t) \approx \hat{U}_{\mathrm{SW}}(t) = \mathcal{T} e^{-i\int_0^t \hat{H}_{\mathrm{SW}}(t') dt' },
\end{align}
to yield the estimate 
\begin{align}
\nonumber
\mathcal{D}(t) &\approx \mathcal{D}_{\mathrm{SW}}(t) = \frac{1}{LS} \sum_{j=1}^L \bra{\Uparrow} \hat{U}_{\mathrm{SW}}^\dagger(t) \hat{S}_j^z \hat{U}_{\mathrm{SW}}(t)\ket{\Uparrow} \\
\label{eq:DSWformula}
&= 1- \frac{1}{LS} \sum_{j=1}^L \bra{\Uparrow} \hat{a}_j^\dagger(t) \hat{a}_j (t) \ket{\Uparrow},
\end{align}
where $\hat{a}_j(t) = \hat{U}_{\mathrm{SW}}^\dagger(t) \hat{a}_j \hat{U}_{\mathrm{SW}}(t)$. We note this expression is easy to compute numerically for large systems (of the order of hundreds of  sites assuming typical present-day computational resources) by 
integrating the Heisenberg-picture time evolution
\begin{align}
\nonumber 
&i[\hat{H}_{\mathrm{SW}}(t),\hat{a}_j] = -i(\eta_{j-1}(t)\hat{a}_{j-1}+ \eta^*_j(t)\hat{a}_{j+1})\\
&-i(\zeta_{j-1}(t)\hat{a}_{j-1}^\dagger + \zeta_j(t)\hat{a}^{\dagger}_{j+1}) - iV_j(t)\hat{a}_j.
\end{align}
Writing the latter in matrix form as 
\begin{equation}
\label{eq:linearsys}
\frac{d}{dt} \begin{pmatrix}
\hat{\vec{a}}(t) \\ \hat{\vec{a}}^\dagger(t)
\end{pmatrix} = -i C(t)\begin{pmatrix}
\hat{\vec{a}}(t) \\ \hat{\vec{a}}^\dagger(t)
\end{pmatrix},
\end{equation}
letting $U(t) = \mathcal{T}e^{-i\int_0^t dt' \, C(t')}$ denote the corresponding time-evolution operator and using the fact that $\hat{a}_j\ket{\Uparrow}=0$, we find that
\begin{equation}
\bra{\Uparrow} \hat{a}_j^\dagger(t) \hat{a}_j(t) \ket{\Uparrow} = \sum_{l=1}^L |U(t)_{j,l+L}|^2.
\end{equation}
Since the matrix $C(t)$ is linear in $S$, the spin-wave prediction for the contrast $\mathcal{D}_{\mathrm{SW}}(t)$ exhibits a scaling collapse in $S$, i.e. there always exists an $S$-independent function $f(t)$ in terms of which 
\begin{equation} 
\mathcal{D}_{\mathrm{SW}}(t) = 1- f(St)/S.
\end{equation}
This observation will prove useful later on when we compare spin-wave theory predictions against numerics.

This is as far as our theory for the contrast $\mathcal{D}(t)$, developed within the time-dependent spin-wave approximation, can be taken in full generality. To make further analytical progress, specific classes of mean-field states and Hamiltonians generating dynamics must be considered.

\section{Perturbations of GZ scars yielding mean-field elliptic traveling wave dynamics}
\label{sec:scars_perturbation}

We now apply our general theory of dynamical instability to the setting of interest in this paper: the GZ scars $|\kappa, q,\gamma\rangle$, which are mean-field states and also exact eigenstates of $\hat{H}_{XYZ}$ with $J_x=\dn(q,\kappa), J_z=\cn(q,\kappa)$, evolving under the perturbed quantum Hamiltonian Eq.~\eqref{eqn:perturbed_H}. 
We shall refer to $\hat{H}_{XYZ}$ with these specific values of $J_x,J_z$ as the ``parent Hamiltonian'' of the GZ scar $|\kappa,q,\gamma\rangle$. 
To remind the reader, the specific perturbations that we choose preserve the form of the Hamiltonian as an XYZ quantum spin chain, and are thus parameterized by a choice of perturbation interaction strengths $(\delta J_x, \delta J_y, \delta J_z)$. Since changing the overall scale of the Hamiltonian is not important, we will maintain the convention that the spin-spin interaction in the $y$-direction is of unit strength even within the perturbed Hamiltonian; thus, we will fix $\delta J_y = 0$ without loss of generality, and restrict the set of perturbations to take the form $(\delta J_x,\delta J_z)$.

In principle, for arbitrary choices of perturbations $(\delta J_x, \delta J_z)$, one has to solve for the new semiclassical trajectory Eq.~\eqref{eq:classEoM} (or Eq.~\eqref{eq:MFEoM}) under $H^\text{cl}$, the classical analog of $\hat{H}$ beginning from the initial state $|\kappa,q,\gamma\rangle$, to obtain the rotation matrix $R_j(t)$. This then yields the rotating frame Hamiltonian $\hat{H}_\text{R}(t)$ and subsequently the spin-wave Hamiltonian $\hat{H}_\text{SW}$, \eqref{eq:SWapprox}. Finally, within the spin-wave approximation, the contrast $\mathcal{D}(t)$ is   given by Eq.~\eqref{eq:DSWformula}. 

While the above prescription is general,  such a computation of $\mathcal{D}(t)$ will generically not be amenable to an analytical treatment. The difficulty stems from the semiclassical trajectory Eq.~\eqref{eq:classEoM} under the perturbed Hamiltonian, which in general will not admit a closed-form expression and can generically only be computed numerically. To make progress, we are thus motivated to ask the following question: for which classes of GZ scars (and their associated parent Hamiltonians) do there exist directions in the $\delta J_x$-$\delta J_z$ space of perturbations such that
  their semiclassical dynamics admits a simple closed-form expression, which might then allow for an analytical understanding of their contrast $\mathcal{D}(t)$?

To this end, we consider the following ansatz of time-dependent mean-field states $|\vec{\bo}(t)\rangle = \otimes_{j=1}^L |\bo_j(t)\rangle$, defined locally by Bloch coherent states $|\bo_j(t)\rangle$ with Bloch vectors
\begin{equation}
\label{eq:GZansatz}
\bo_j = \begin{pmatrix} \alpha\, \mathrm{cn}(qj-\omega t,\kappa) \\
\beta\, \mathrm{sn}(qj-\omega t,\kappa) \\
\gamma\, \mathrm{dn}(qj-\omega t,\kappa)\end{pmatrix}
\end{equation}
for some $\omega$, which represents an 
`elliptic traveling wave' ansatz~\cite{granovskii1985periodic,roberts1988dynamics,lakshmanan2008dynamic}. Above, $\alpha,\beta,\gamma$ are as in Eq.~\eqref{eqn:scar_state}.
Note that at time $t = 0$, this dynamical mean-field state reduces to the GZ scar:
\begin{align}
 |\vec{\bo}(0)\rangle = |\kappa,q,\gamma\rangle,
\end{align}
thus satisfying the desired initial conditions. 
We now ask which perturbed Hamiltonians have the elliptic traveling wave ansatz Eq.~\eqref{eq:GZansatz} as a solution of their discrete Landau-Liftshitz equations of motion in the semiclassical limit. 

Substituting the ansatz Eq.~\eqref{eq:GZansatz} into Eq.~\eqref{eq:classEoM}, we find (temporarily reinstating $\delta J_y$ but remembering that we will eventually set it to $0$) that they are solutions if and only if 
\begin{align}
\label{eq:MF1}
 \frac{2 S\beta \gamma\left( \delta J_y \cn(q,\kappa) - \delta J_z \right)\dn(q,\kappa)}{1-\kappa^2 \sn^2(u_j,\kappa)\sn^2(q,\kappa)} &= \alpha \omega,  \\ 
 \label{eq:MF2}
 \frac{2 S\alpha \gamma\left(    \delta J_x \mathrm{cn}(q,\kappa) -  \delta J_z \mathrm{dn}(q,\kappa)\right)}{1-\kappa^2 \sn^2(u_j,\kappa)\sn^2(q,\kappa)} &= \beta \omega, \\ 
 \label{eq:MF3}
 \frac{2S \alpha \beta\left(  \delta J_x - \delta J_y \mathrm{dn}(q,\kappa)\right)}{1-\kappa^2 \sn^2(u_j,\kappa)\sn^2(q,\kappa)} &= \gamma \kappa^2 \omega,
\end{align}
where $u_j := qj -\omega t$.

We identify (non-exhaustively) three simple classes of 
solutions to Eqs.~\eqref{eq:MF1}, \eqref{eq:MF2}, \eqref{eq:MF3}:

{\it Transverse spin-helices}. --- 
First,  set $\kappa = 0$ and parameterize $\gamma = \cos\theta$ for some $\theta \in [0,\pi]$, which amounts to the parent Hamiltonian of the GZ scar $|0,q,\cos\theta\rangle$ being of XXZ type with $J_x = 1$ and anisotropy $J_z = \cos q$.   Then, the ansatz Eq.~\eqref{eq:GZansatz} reduces to the familiar~\cite{popkov2021phantom,jepsen2021transverse,Jepsen_2022,popkov2023universality} `rotating spin helix' 
\begin{equation}
\label{eq:tsh}
\bo_j(t) = \begin{pmatrix}
  \sin{\theta}\cos(qj-\omega t) \\ \sin{\theta}\sin{(qj-\omega t)} \\ \cos{\theta}
\end{pmatrix},
\end{equation}
(see Fig.~\ref{fig:GZprofiles}, left), 
and we find that it is a solution to the Landau-Liftshitz equations if the perturbations are of the form $(0,0,\delta J_z)$ for arbitrary $\delta J_z$, i.e., perturbations of the spin-spin interactions along the $z$-direction,  
and the
 rotation frequency 
\begin{equation}
\label{eq:helixfreq}
\omega = -2S\cos{\theta} \delta J_z.
\end{equation}


We refer to this class of mean-field states as ``transverse spin helices'', following~\cite{popkov2021phantom,jepsen2021transverse,Jepsen_2022,popkov2023universality}.
This is the most analytically tractable case that we study in this paper, as well as enjoying immediate relevance to recent ultracold atom experiments~\cite{Jepsen_2020,jepsen2021transverse,Jepsen_2022}. 
\\

{\it Generalized transverse spin helices}. --- Next, assume $0 < \kappa <1$. Then, the denominators of the left hand side of Eqs.~\eqref{eq:MF1}, \eqref{eq:MF2}, \eqref{eq:MF3} are spatially and temporally inhomogeneous in general. In order for there to be a solution,  the numerators should vanish, which means the frequency $\omega = 0$. 
Eqs.~\eqref{eq:MF1}, \eqref{eq:MF2}, \eqref{eq:MF3} reduce to~\footnote{Note these equations were solved by Granovskii and Zhedanov~\cite{granovskii1985periodic} and independently by Roberts and Tompson~\cite{roberts1988dynamics}.}
\begin{align}
\label{eq:GZ1}
2 \beta \gamma (\delta J_y \cn(q,\kappa) - \delta J_z)\dn(q,\kappa) &= 0,  \\ 
\label{eq:GZ2}
2 \alpha \gamma (\delta J_x \mathrm{cn}(q,\kappa) - \delta J_z\mathrm{dn}(q,\kappa)) &= 0,  \\ 
\label{eq:GZ3}
2 \alpha \beta (\delta J_x - \delta J_y\mathrm{dn}(q,\kappa)) &= 0.
\end{align}

If we set $\gamma = 0$,  which yields Bloch vectors defining the spin texture 
\begin{equation}
\label{eq:gtsh}
\bo_j = \begin{pmatrix}
  \cn(qj,\kappa) \\ \sn(qj,\kappa) \\ 0
\end{pmatrix},
\end{equation}
then Eqs.~\eqref{eq:GZ1}, \eqref{eq:GZ2} are automatically satisfied, while Eq.~\eqref{eq:GZ3} implies that perturbations can be chosen of the form $(0, 0,\delta J_z)$ for arbitrary $\delta J_z$. In other words, the GZ scar $|\kappa,q,0\rangle$  associated with Eq.~\eqref{eq:gtsh} with parent XYZ Hamiltonian defined by $J_x = \dn(q,\kappa)$ and $J_z = \cn(q,\kappa)$ remains a {\it static} solution of the classical Landau-Lifshitz equations for perturbations involving spin-spin interactions along the $x$-direction, though it is no longer a quantum eigenstate. 

Just like the transverse spin helices, Eq.~\eqref{eq:gtsh} describes a spin texture winding in the $x$-$y$ plane but in a more complicated fashion (see Fig.~\ref{fig:GZprofiles}, middle); we will hence refer to these states as ``generalized transverse spin helices''. 
\\

{\it Generalized longitudinal spin helices}. --- 
For $0 < \kappa <1$, we find there is another way for Eqs.~\eqref{eq:GZ1}, \eqref{eq:GZ2}, \eqref{eq:GZ3} to be satisfied. We set $\gamma = 1$, which define Bloch vectors 
\begin{equation}
\label{eq:glsh}
\bo_j = \begin{pmatrix}
0 \\ \kappa \sn(qj,\kappa) \\ \dn(qj,\kappa)
\end{pmatrix}.
\end{equation}
This describes a spin texture winding in the $y$-$z$ plane~\cite{granovskii1985periodic,roberts1988dynamics,lakshmanan2008dynamic}, hence we refer to such states as ``generalized longitudinal spin helices'' (see Fig.~\ref{fig:GZprofiles}, right). 

We find these are static solutions of the Landau-Lifshitz equations for perturbations involving spin-spin interactions along the $x$-direction, that is, perturbations of the parent Hamiltonian of the form $(\delta J_x,0,0)$ for arbitrary $\delta J_x$. Note that as $\kappa \to 0^+$, Eq.~\eqref{eq:glsh} reduces to the ferromagnetic $z$-polarized state in the XXZ Hamiltonian, while as $\kappa \to 1^-$ it becomes a domain wall configuration (i.e., the wavelength of winding becomes infinite).

\section{Dynamical instability of transverse spin helices in spin-$S$ XXZ chains}
\label{sec:TSH}
We now derive the predictions arising from our general theory of dynamical instability, applied to the three classes of GZ scars and the perturbations of their parent Hamiltonians found in Sec.~\ref{sec:scars_perturbation}. 

We start with the transverse spin helices $|0,q,\cos\theta\rangle$, assuming $ 0 < q < \pi/2$ and $0 < \theta < \pi$, which we remind the reader are GZ quantum many-body scars of the spin-$S$ XXZ Hamiltonian 
\begin{align}
\hat{H}_{XXZ} =  \sum_{j=1}^L   \hat{S}_j^x \hat{S}_{j+1}^x +  \hat{S}_j^y \hat{S}_{j+1}^y + (\cos q) \hat{S}_j^z \hat{S}_{j+1}^z.
\end{align}
As discussed above, we perturb the XXZ Hamiltonian with the term $\delta \hat{H} = \delta J_z\sum_{j=1}^L   \hat{S}^z_j \hat{S}^z_{j+1}$ for some $\delta J_z \in \mathbb{R}$. The transverse spin helix $|0,q,\cos\theta\rangle$ for $\delta J_z \neq 0$  will   no longer be an eigenstate and will evolve in time. 


\subsection{Bogoliubov theory of linear stability}
Following Sec.~\ref{sec:gentheory}, we work out in Appendix   \ref{app:rotframe}   the spin-wave Hamiltonian Eq.~\eqref{eq:HSW} and find that it is translation invariant,
\begin{align}
\nonumber 
\hat{H}_{\mathrm{SW}} = &\sum_{j=1}^L\eta \hat{a}_{j+1}^\dagger\hat{a}_j +\eta^*\hat{a}_j^\dagger \hat{a}_{j+1} + \zeta (\hat{a}_j^\dagger \hat{a}_{j+1}^\dagger + \hat{a}_{j+1} \hat{a}_j) \\
&+  V \hat{a}_j^\dagger \hat{a}_j,
\end{align}
where $\eta = (S/2)\left(2\cos{q} + \sin^2{\theta}\,\delta J_z-i 2 \cos{\theta}\sin{q}\right)$, $
\zeta = (S/2)\sin^2{\theta}\,\delta J_z$ and $V = -2S\cos{q}$.

\begin{figure}[t]
    \centering
\includegraphics[width=0.99\linewidth]{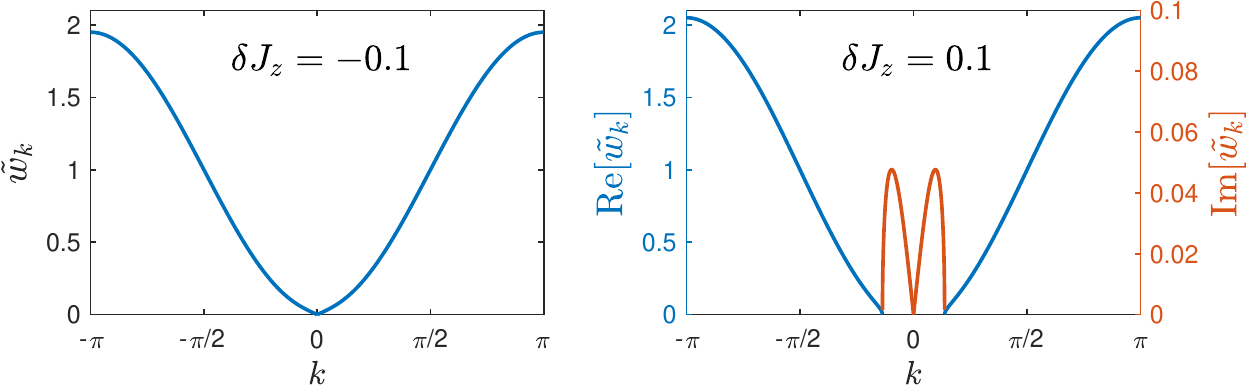}
    \caption{Illustrative plots of the dispersion relation $\tilde{w}_k$  for perturbations  giving rise to ({\bf Left}) the stable regime ($\delta J_z < 0$) and ({\bf Right}) the unstable regime ($\delta J_z > 0$), for $q=\pi/3$, $\theta=\pi/4$. 
    In the stable regime, the dispersion is purely real, while in the unstable regime, there is a non-zero range of quasimomenta $k$ where the dispersion is purely imaginary.}
    \label{fig::DecayFactor}
\end{figure}

After a standard combined Fourier and Bogoliubov transformation to bosonic modes $\hat{b}_k$ with quasimomentum $k$, we obtain the diagonalized Bogoliubov form of the spin-wave Hamiltonian
\begin{equation}
\label{eq:SWHamiltonian}
\hat{H}_{\mathrm{SW}} = \sum_{k \in \mathrm{BZ}} \omega^{\mathrm{SW}}_k \hat{b}_k^\dagger \hat{b}_k, 
\end{equation}
which has dispersion relation
\begin{align}
\omega^{\mathrm{SW}}_k = \frac{B_k^- + \mathrm{sign}(B_k^+) \sqrt{(B_k^+)^2 - 4A_k^2}}{2}, 
\label{eq:swdisp}
\end{align}
where $A_k = S\sin^2{\theta} \cos{k}\,\delta J_z$,  $B_k^\pm = B_k\pm B_{-k}$,
$B_k = S\big(\sin^2{\theta}\cos{k}\,\delta J_z - 4\cos{q}\sin^2{(k/2)}
-2\cos{\theta}\sin{q}\sin{k}\big)$ and $\mathrm{BZ}$ denotes the first Brillouin zone. Importantly, whenever $ -\frac{\cos{q}}{\sin^2{\theta}} \leq \delta J_z \leq 0$, which we call the `stable' regime, the single-particle spectrum is real for all $k$, while for $\delta J_z > 0$, which we call the `unstable' regime, it is  {\it imaginary} for some $k$. 
In Fig.\,\ref{fig::DecayFactor} we plot a related dispersion capturing the same stable-unstable behavior \footnote{They capture the same behavior since  $\mathrm{Im}[\tilde{w}_k] \neq 0$ whenever $\mathrm{Im}[\omega_k^{\mathrm{SW}}] \neq 0$.},
\begin{align}
\tilde{w}_k & = \frac{1}{2S}\text{sign}(B_k^+)(\omega_k^\text{SW} +\omega_{-k}^\text{SW}) \nonumber \\
& = 2\sqrt{2}  \sin{(k/2)}\sqrt{\cos{q}} \nonumber \\
& \times \sqrt{2\cos{q}\sin^2{(k/2)}-\sin^2{\theta}\cos{k}\,\delta J_z},
\end{align}
and which will be be relevant for the expression for the contrast later in Eq.\,\ref{eqn:f_scaling}.
%


The presence of imaginarity in the dispersions $\omega_k^\text{SW}$ or $\tilde{w}_k$ implies an exponential growth of the associated modes in time, i.e., an instability.
Thus, time-dependent spin-wave theory would predict the existence of an {\it asymmetric} instability about the point $\delta J_z = 0$ for sufficiently large spin $S$, as we shall see explicitly in the analysis of the contrast below.

\subsection{Dynamics of the contrast}

\begin{figure*}[t]
    \centering
    \includegraphics[width=0.95\linewidth]{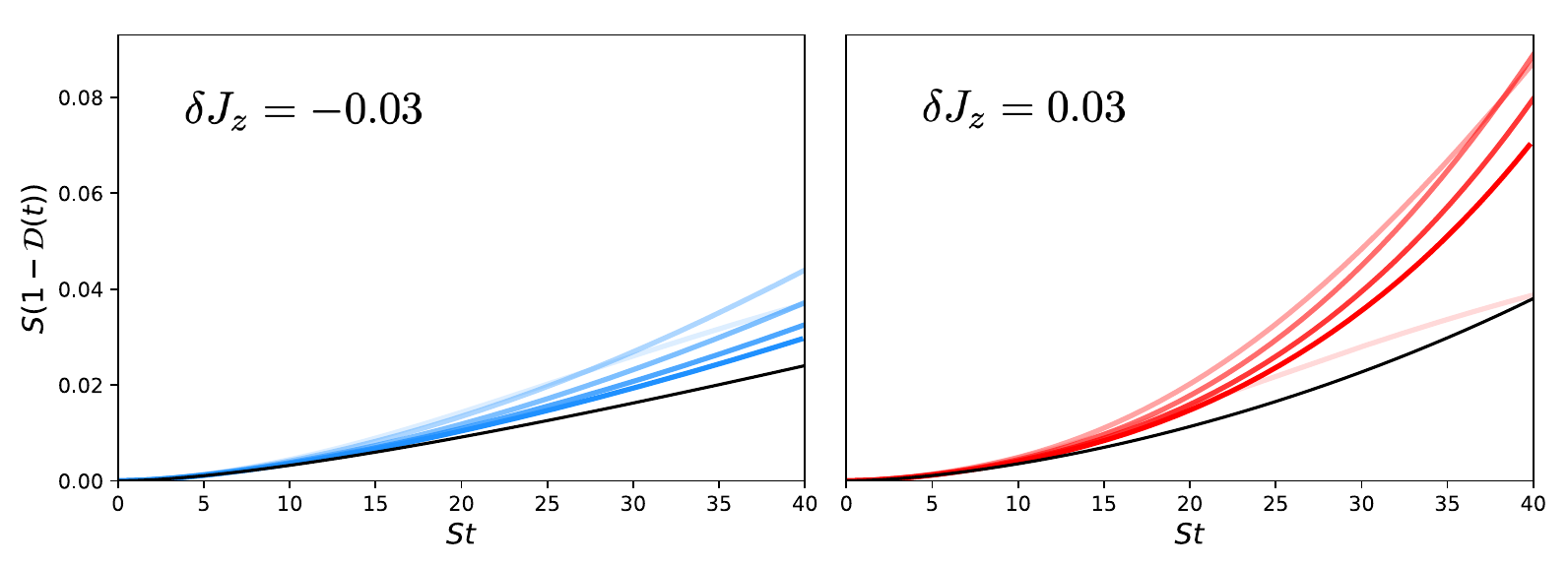}
    \caption{Decay of the contrast $\mathcal{D}(t)$  between the stable ({\bf Left}) and unstable regimes ({\bf Right}) on perturbing transverse spin-helix states in spin-$S$ XXZ chains with initial anisotropy $J_z = 0.5$. The initial spin texture at each site is given by Eq.~\eqref{eq:tsh} with $q = \pi /3$ and $\theta=\pi/4$. We plot the empirical scaling function $S(1-\mathcal{D}(t))$ for iTEBD data as a function of $St$ and compare it to spin-wave theory predictions Eq. \eqref{eqn:f_scaling} (black lines). We set the detuning parameter to $\delta J_z = -0.03$ (stable regime, left plot) and $\delta J_z = 0.03$ (unstable regime, right plot). Increasing opacity corresponds to the spin $S$ increasing from $S=1/2$ to $S=5/2$ in half-integer increments. It is clear that the contrast decays faster in the unstable regime than in the stable regime, despite both plots exhibiting the same magnitude of detuning $|\delta J_z|=0.03$. The dynamics for $S=1/2$ is an outlier that we attribute to microscopic integrability at this special value of $S$. For $S \geq 1$ (faintest curve), the numerical data converges to the spin-wave theory prediction as $S$ increases, in a manner consistent with asymmetric decay as a function of the detuning parameter $\delta J_z$.
    } 
    \label{fig:kappa0numerics}
\end{figure*}

Previous experimental realizations of spin helices in the XXZ chain~\cite{Jepsen_2022} diagnosed the decay of these helices via the `spin contrast', which in the thermodynamic limit is given by 
\begin{equation}
\mathcal{C}(t)  = \lim_{L \to \infty}\frac{2}{LS\sin{\theta}}\sum_{j=1}^{L} \langle \psi(t) | \hat{S}_j^x |\psi(t)\rangle \cos{(qj-\omega t)},
\label{eqn:spin-contrast}
\end{equation}
with $\omega$ as in Eq.~\eqref{eq:helixfreq}.

This is related to our more general notion of contrast Eq.~\eqref{eq:defcontrast} by the relation
\begin{equation}
\mathcal{C}(t) = \frac{\mathcal{D}(t) - \cos^2{\theta}}{\sin^{2}{\theta}},
\end{equation}
which is not hard to show from the definitions of the rotation matrices given in Appendix \ref{app:rotframe}. Thus the dynamics of $\mathcal{C}(t)$ is equivalent to the dynamics $\mathcal{D}(t)$, and it suffices to focus on the latter quantity.

In the thermodynamic limit, the spin-wave theory formula Eq.~\eqref{eq:DSWformula} for the contrast can be expressed as an integral
\begin{align}
\nonumber
&\frac{1}{LS} \sum_{j=1}^L \bra{\Uparrow} \hat{a}^\dagger_j(t)\hat{a}_j(t)\ket{\Uparrow} \\
\to &\frac{1}{2\pi S} \int_{-\pi}^{\pi} dk \, \left(\frac{A_k^2}{\tilde{w}_k^2S^2}\right) \sin^2{(\tilde{w}_k St)}, \quad L \to \infty.
\end{align}
Thus, a large-$S$ approximation to the dynamics of the spin contrast is given by
\begin{equation}
\mathcal{D}(t)  \approx \mathcal{D}_{\mathrm{SW}}(t) = 1-f(St)/S \end{equation}
where the scaling function
\begin{equation}
f(\tau) = \frac{1}{2\pi} \int_{-\pi}^{\pi} dk \, \left(\frac{\tilde{A}_k^2}{\tilde{w}_k^2}\right) \sin^2{(\tilde{w}_k\tau)}.
\label{eqn:f_scaling}
\end{equation}
is independent of $S$, since $\tilde{A}_k := A_k/S$ and $\tilde{w}_k$ are $S$-independent functions.

\subsection{Prediction of asymmetric decay}
Since $\mathrm{Im}[\tilde{w}_k] \neq 0$ whenever $\mathrm{Im}[\omega_k^{\mathrm{SW}}] \neq 0$, the behavior of $f(\tau)$ at late times depends crucially on whether the quench is to the stable or to the unstable regime. In the stable regime, we expect from Eq.~\eqref{eqn:f_scaling} that the contrast decays solely by dephasing from interference between fluctuations about the quantum many-body scar. In the unstable regime, the contrast should be dominated by the (exponential) growth of the most unstable mode. 

In Appendix \ref{app:asymp}, we  rigorously extract the leading asymptotic behavior at late times, finding
\begin{equation}
\label{eq:mainresult}
\mathcal{D}_{\mathrm{SW}}(t) \sim \begin{cases} - \gamma_1 t, &  -\frac{\cos{q}}{\sin^2{\theta}} < \delta J_z < 0\\ -\frac{A}{S\sqrt{St}}e^{\gamma_2 t}, &  \hspace{1.1cm} \delta J_z >0
\end{cases}
\end{equation}
as $t \to \infty$, where $A$ is a non-universal, $S$-independent constant, and the decay rates
\begin{equation}
\label{eq:gamma1}
\gamma_1 = \frac{1}{2\sqrt{2}} \frac{\sin^3{\theta}}{\sqrt{\cos{q}}} |\delta J_z|^{3/2}
\end{equation}
and 
\begin{equation}
\label{eq:gamma2}
\gamma_2 \approx 2 S (\sin^2{\theta}) \delta J_z, \quad 0 < \delta J_z \ll 1
\end{equation}
in the perturbatively unstable regime.

Eqs.~\eqref{eq:mainresult}, \eqref{eq:gamma1} and \eqref{eq:gamma2} are our main results for the transverse spin helices: they predict a dramatic asymmetry in both the temporal profiles and the decay rate of the spin contrast at long times, which becomes exact in the semiclassical limit $S \to \infty$.
Namely, in the stable regime, the dynamics is slow and linear; while in the unstable regime, it is fast and exponential. 
We stress our results' inherently non-perturbative nature: our predictions (in particular, the non-analyticity of the rate Eq.~\eqref{eq:gamma1})  lie beyond all perturbative approaches to relaxation in many-body systems that we are aware of such as short-time expansions or Fermi's Golden Rule, the latter of which, as argued before, would predict a decay that is  insensitive to the sign of the perturbation and an algebraic rate $\propto (\delta J_z)^2$, in contradiction with our results.

\subsection{Numerical simulations}
To verify that our predictions hold more generally away from large $S$, we perform tensor-network numerical simulations of spin-helix relaxation from the XXZ Hamiltonian, namely, matrix-product state simulations (MPS) with the infinite time-evolving block decimation (iTEBD) method (details in Appendix \ref{app:numerical_cal}) for numerically accessible values of small $S$, choosing $J_z = 0.5$ and a spin-helix scar with $q = \pi/3, \theta = \pi/4$ and then quenching the anisotropy to generate dynamics. We compute $\mathcal{D}(t)$ for  different $S$ and form the `empirical scaling function' $f_{\mathrm{emp}.}(\tau) = S(1-\mathcal{D}(\tau/S))$, which we then compare to the analytical prediction $f(\tau)$ from spin-wave theory, see Fig.~\ref{fig:kappa0numerics}. We expect that $f_{\mathrm{emp.}}(\tau) \to f(\tau)$
pointwise as $S \to \infty$. Indeed, we observe evidence for such convergence in Fig. \ref{fig:kappa0numerics} in both the stable and unstable regimes as $S$ increases. Moreover, already for small values of $S$ ($S = 1,3/2,2,5/2$)~\footnote{$S=1/2$ is an exception. This is expected because the model with $S=1/2$ is quantum integrable~\cite{korepin1997quantum}, in contrast to $S>1/2$, and should therefore exhibit qualitatively different relaxation dynamics.} 
agreement between the numerics and the limiting spin-wave prediction is apparent, which demonstrates in particular that our prediction of asymmetry in the decay profiles of the spin-helix scar is applicable far from the semiclassical limit.

\section{Dynamical instability of generalized spin helices in spin-$S$ XYZ chains}
\label{sec:GSH}

Next, we move to the generalized transverse spin helices $|\kappa,q,0\rangle$ and generalized transverse longitudinal helices  $|\kappa,q,1\rangle$,  assuming $0 < \kappa < 1$ and $0 < q < K(\kappa)$, both of which are GZ quantum many-body scars of the spin-$S$ XYZ Hamiltonian 
\begin{align}
\hat{H}_{XYZ} =  \sum_{j=1}^L  \dn(q,\kappa) \hat{S}_j^x \hat{S}_{j+1}^x +  \hat{S}_j^y \hat{S}_{j+1}^y + \cn(q,\kappa) \hat{S}_j^z \hat{S}_{j+1}^z.
\end{align}
For the former, we perturb the XYZ Hamiltonian with the term $\delta \hat{H} = \delta J_z\sum_{j=1}^L   \hat{S}^z_j \hat{S}^z_{j+1}$ for some $\delta J_z \in \mathbb{R}$. 
For the latter, we perturb the XYZ Hamiltonian with the term $\delta \hat{H} = \delta J_x\sum_{j=1}^L   \hat{S}^x_j \hat{S}^x_{j+1}$ for some $\delta J_x\in \mathbb{R}$.

  We work out in Appendix   \ref{app:rotframe} the spin-wave Hamiltonians governing these scars' linear stability. 
  Remarkably, they have the same form in both cases: 
  the spin-wave Hamiltonian has translation invariant hopping and pair-creation terms while the potential term is spatially-dependent, namely
\begin{align}
\nonumber 
\hat{H}_{\mathrm{SW}} = &\sum_{j=1}^L\eta (\hat{a}_{j+1}^\dagger\hat{a}_j +\hat{a}_j^\dagger \hat{a}_{j+1}) + \zeta (\hat{a}_j^\dagger \hat{a}_{j+1}^\dagger + \hat{a}_{j+1} \hat{a}_j) \\
\label{eq:gshHSW}
&+  V_j \hat{a}_j^\dagger \hat{a}_j,
\end{align}
where in the transverse case, $\eta=(S/2)(2\cn(q,\kappa)+\delta J_z)$, $\zeta=(S/2)\delta J_z$ and
\begin{equation}
\label{eq:onsiteV}
V_j = -\frac{2S\cn(q,\kappa)\dn(q,\kappa)}{1-\kappa^2 \sn^2(qj,\kappa)\sn^2(q,\kappa)}
\end{equation}
while in the longitudinal case $\eta=(S/2)(2\dn(q,\kappa)+\delta J_x)$, $\zeta=(S/2)\delta J_x$ and $V_j$ is unchanged from Eq.~\eqref{eq:onsiteV}.

Thus in both cases, the spin-wave theory reduces to the theory of a free boson hopping in a periodic potential, with wavelength
\begin{equation}
\lambda = 4K(\kappa)/q = L/M.
\end{equation}
The quadratic spin-wave Hamiltonian Eq.~\eqref{eq:gshHSW} can either be diagonalized in finite size by an elementary, brute-force diagonalization of the linear Heisenberg-picture dynamics Eq.~\eqref{eq:linearsys} or by using standard techniques of spin-wave theory, i.e., a Fourier transform followed by a Bogoliubov transform. The latter is tractable in the thermodynamic limit and will therefore be pursued here.

However, unlike for more conventional applications of Bogoliubov theory, the unit cell of the Hamiltonian Eq.~\eqref{eq:gshHSW} consists of $\lambda$ basis sites and can be order one or extensively large depending on the value of $M$. Put differently, the number of flavours of spin waves that needs to be considered in the Bogoliubov theory is determined by the period of the Jacobi elliptic function $\sn(x,\kappa)$ via Eq.~\eqref{eq:onsiteV}, and therefore depends sensitively on our choice of parameters.

\subsection{Multi-flavour Bogoliubov theory}
\begin{figure}[t]
    \centering
\includegraphics[width=0.99\linewidth]{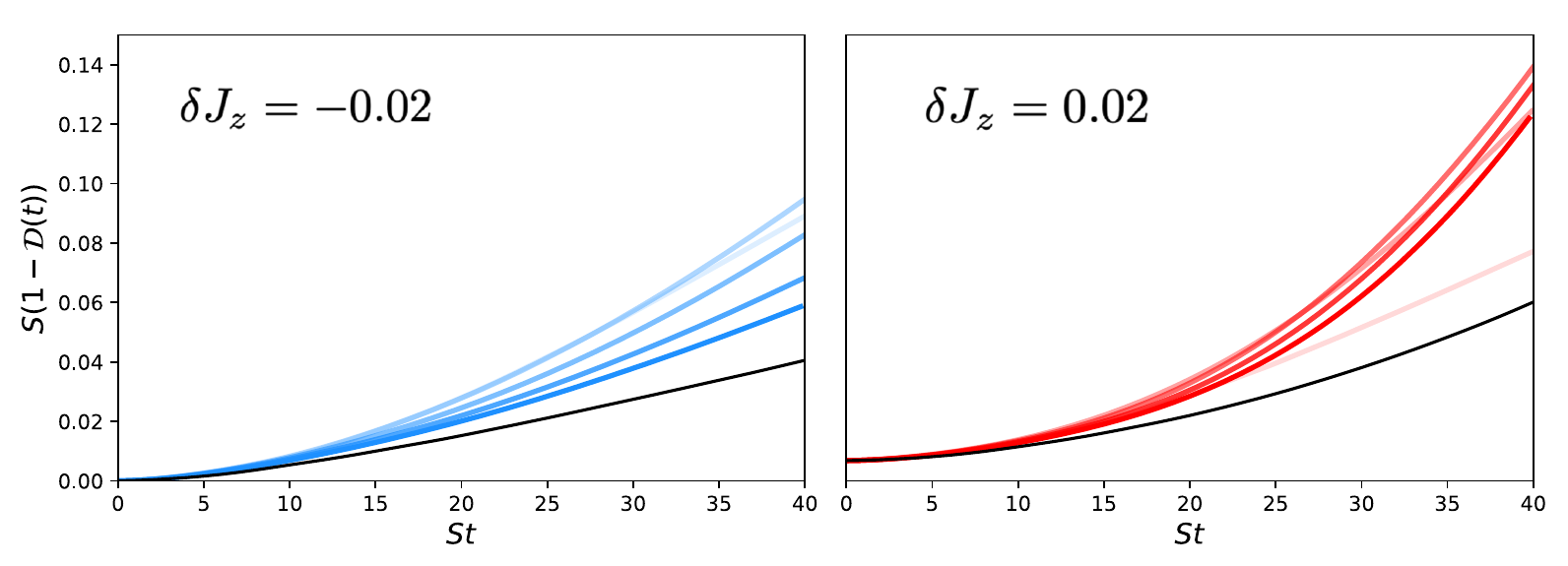}
    \caption{Asymmetric decay of generalized transverse spin helices in spin-$S$ XYZ chains as captured by the contrast $\mathcal{D}(t)$. The initial spin texture is given by Eq.~\eqref{eq:gtsh} with $\kappa = 0.9$ and $q = 2K(\kappa)/3$, corresponding to a wavelength of $\lambda=6$ sites. This GZ scar is an exact product eigenstate when the XYZ couplings $J_x = 0.537...$, $J_y=1$ and $J_z = 0.349...$ but decays asymmetrically in perturbation strength $\delta J_z$. 
    We simulate the resulting dynamics of the contrast $\mathcal{D}(t)$ for detuning parameters $\delta J_z = -0.02$ (stable regime, left plot) and $\delta J_z = 0.02$ (unstable regime, right plot), with plotting conventions as in Fig.~\ref{fig:kappa0numerics} except that analytical predictions (black lines) are now obtained from numerically integrating the multi-flavour Bogoliubov theory prediction Eq.~\eqref{eq:gshcontrast}.}
    \label{fig:transhelixnumerics}
\end{figure}
To derive the Bogoliubov theory of Eq.~\eqref{eq:gshHSW}, we treat the $L$ sites of the physical lattice as $M$ unit cells consisting of $\lambda$ basis sites each (thus it is simplest to assume that $L = \lambda M$ with $\lambda, M$ integers). We also relabel the bosonic operators to reflect this unit cell structure, as $\hat{a}_{l,\sigma} = \hat{a}_{(l-1)\lambda+\sigma}$ where the unit-cell index $l=1,\ldots,M$, the `flavour' or basis-site index $\sigma = 1,2,\ldots \lambda$ and periodic boundary conditions set $\hat{a}_{M+1,\sigma}\equiv \hat{a}_{1,\sigma}$. Defining Fourier transformed variables $\hat{a}_{k,\sigma} =\frac{1}{\sqrt{M}} \sum_{l=1}^M e^{-ikl} \hat{a}_{l,\sigma} $ on this sublattice and writing $\mathrm{BZ}'$ for the sublattice Brillouin zone, we can write the Hamiltonian as a sum over momentum and flavour indices
\begin{align}
\nonumber
\hat{H}_{\mathrm{SW}} = \sum_{k\in\mathrm{BZ}'} \sum_{\sigma,\sigma'=1}^\lambda &\frac{1}{2}(A_{k,\sigma,\sigma'} \hat{a}^\dagger_{k,\sigma} \hat{a}^\dagger_{-k,\sigma'} +A_{k,\sigma,\sigma'}^*  \hat{a}_{-k,\sigma'}\hat{a}_{k,\sigma}) \\ &+ B_{k,\sigma,\sigma'} \hat{a}_{k,\sigma}^\dagger \hat{a}_{k,\sigma'}
\label{eq:gshmomspaceH}
\end{align}
where $A_k$ and $B_k$ are square, Hermitian $\lambda$-by-$\lambda$ matrices given explicitly by
\begin{equation}
A_k =
\begin{pmatrix} 
0 & \zeta & 0 & \ldots & e^{-ik}\zeta \\ 
\zeta & 0 & \zeta & \ldots  & 0 \\ 
 \vdots & \ddots & \ddots & \ddots  & \vdots \\
0 & \ldots &  \zeta & 0  & \zeta \\ 
e^{ik} \zeta & 0 & \ldots &\zeta & 0
\end{pmatrix}
\end{equation}
and
\begin{equation}
B_k = 
\begin{pmatrix} 
V_1 & \eta & 0 & \ldots & e^{-ik}\eta \\ 
\eta & V_2 & \eta & \ldots  & 0 \\ 
 \vdots & \ddots & \ddots & \ddots  & \vdots \\
0 & \ldots &  \eta & V_{\lambda-1}  & \eta \\ 
e^{ik} \eta & 0 & \ldots &\eta & V_{\lambda}
\end{pmatrix}
\end{equation}
for all $k \in \mathrm{BZ}'$. Then the Heisenberg equations of motion for the Hamiltonian Eq. \eqref{eq:gshmomspaceH} can be written as
\begin{equation}
\label{eq:gshmatrixevo}
\frac{d}{dt} \begin{pmatrix} \hat{\vec{a}}_k(t) \\ \hat{\vec{a}}_{-k}^\dagger(t) \end{pmatrix} = -i C_k \begin{pmatrix}
\hat{\vec{a}}_k(t) \\ \hat{\vec{a}}_{-k}^\dagger(t)
\end{pmatrix}
\end{equation}
where the $2\lambda$-by-$2\lambda$ matrices $C_k = 
\begin{pmatrix} B_k & A_k \\ -A_k & -B_{k} \end{pmatrix}$ and we write $\hat{a}_k(t) = \hat{U}_{\mathrm{SW}}^\dagger(t) \hat{a}_k \hat{U}_{\mathrm{SW}}(t)$. From Eq.~\eqref{eq:gshmatrixevo} we can both deduce the dynamics of the contrast Eq.~\eqref{eq:DSWformula} as in Section \ref{sec:gentheory} and determine when perturbed generalized spin-helix states become linearly unstable. 

We first consider the dynamics of the contrast. To this end, define the evolution operator $U(t) = e^{-iC_kt}$ for Eq.~\eqref{eq:gshmatrixevo} and note that $
 \sum_{j=1}^L \bra{\Uparrow} \hat{a}_j^\dagger(t) \hat{a}_j(t) \ket{\Uparrow}
 = \sum_{k \in \mathrm{BZ}'}\sum_{\sigma=1}^\lambda \bra{\Uparrow} \hat{a}_{k,\sigma}^\dagger(t) \hat{a}_{k,\sigma}(t) \ket{\Uparrow}$. Then since $\hat{a}_k \ket{\Uparrow} = 0$, Eq.~\eqref{eq:gshmatrixevo} implies that
\begin{equation}
\bra{\Uparrow} \hat{a}_{k,\sigma}^\dagger(t)\hat{a}_{k,\sigma}(t)\ket{\Uparrow} = \sum_{\sigma'=1}^{\lambda} |U(t)_{k,\sigma,\lambda+\sigma'}|^2
\end{equation}
and it follows by Eq.~\eqref{eq:DSWformula} that the spin-wave theory approximation to the contrast reads
\begin{align}
\nonumber
\mathcal{D}_{\mathrm{SW}}(t) &= 1 - \frac{1}{LS} \sum_{k \in \mathrm{BZ}'} \sum_{\sigma,\sigma'=1}^{\lambda}|U(t)_{k,\sigma,\lambda+\sigma'}|^2 \\
&\to 1 - \frac{1}{2\pi \lambda S} \sum_{\sigma,\sigma'=1}^{\lambda} \int_{-\pi}^{\pi} dk \, |U(t)_{k,\sigma,\lambda+\sigma'}|^2
\label{eq:gshcontrast}
\end{align}
in the thermodynamic limit $L \to \infty$.

To determine the long-time stability of solutions to Eq.~\eqref{eq:gshmatrixevo}, it is simplest to introduce canonical and Hermitian operators $\hat{q}_k = (\hat{a}_k+\hat{a}_k^\dagger)/\sqrt{2}$ and $\hat{p}_k=(\hat{a}_k-\hat{a}_k^\dagger)/(i\sqrt{2})$ and write $\hat{\vec{x}}_k = (\hat{\vec{q}}_k, \hat{\vec{q}}_{-k}, \hat{\vec{p}}_k, \hat{\vec{p}}_{-k})$ for $k \neq 0,\pi$ and $\hat{\vec{x}}_k = (\hat{\vec{q}}_k, \hat{\vec{p}}_k)$ when $k = 0, \pi$ (note that the case $k=\pi$ only arises when $M$ is even). Then Eq.~\eqref{eq:gshmatrixevo} implies the canonical equations of motion
\begin{equation}
\label{eq:gshcanonical}
\frac{d}{dt}\vec{x}_k(t) = D_k \vec{x}_k(t)
\end{equation}
where we have defined the dynamical matrix
\begin{equation}
D_k = \begin{pmatrix} \mathrm{Im}[B_k] & \mathrm{Im}[A_k] & \mathrm{Re}[B_k] & -\mathrm{Re}[A_k] \\
-\mathrm{Im}[A_k] & -\mathrm{Im}[B_k] & -\mathrm{Re}[A_k] & \mathrm{Re}[B_k] \\
-\mathrm{Re}[B_k] & -\mathrm{Re}[A_k] & \mathrm{Im}[B_k] & -\mathrm{Im}[A_k]\\
-\mathrm{Re}[A_k] & -\mathrm{Re}[B_k] & \mathrm{Im}[A_k] & -\mathrm{Im}[B_k]
\end{pmatrix}
\label{eqn:Dk1}
\end{equation}
which is a $4\lambda$-by-$4\lambda$ matrix for $k \neq 0,\pi$, and
\begin{equation}
D_k = \begin{pmatrix} \mathrm{Im}[B_k+A_k] & \mathrm{Re}[B_k-A_k] \\ -\mathrm{Re}[B_k+A_k] & \mathrm{Im}[B_k-A_k]
\label{eqn:Dk2}
\end{pmatrix},
\end{equation}
which is a $2\lambda$-by-$2\lambda$ matrix, for $k = 0, \pi$. It is clear by reflection symmetry of the eigenvalues of $D_k$ in the imaginary axis~\cite{kapitula2013spectral} that a given generalized spin-helix state will be linearly stable iff the eigenvalues of $D_k$ are purely imaginary for all $k \in \mathrm{BZ}'$. 

\subsection{Generalized transverse spin-helix states}
\label{sec:GTSH}

We find that numerically diagonalizing the 
dynamical matrices Eqs.~\eqref{eqn:Dk1}, \eqref{eqn:Dk2} 
for generalized transverse helices with wavelengths exceeding four sites (i.e., for those $q$ satisfying $0<q<K(\kappa)$) predicts that just as for conventional transverse spin helices, there is a neighborhood of $\delta J_z = 0$ in which the generalized transverse spin helix will always be stable when $\delta J_z < 0$ and unstable when $\delta J_z > 0$, i.e., an asymmetric instability about $\delta J_z = 0$. 

We corroborate our predictions against microscopic iTEBD simulations in Fig.~\ref{fig:transhelixnumerics}, again seeing evidence for asymmetric decay and partial convergence to the spin-wave theory prediction Eq.~\eqref{eq:gshcontrast} as the spin $S = 1/2, 1,3/2,2,5/2$ increases, far from the semiclassical limit.

\subsection{Generalized longitudinal spin-helix states}
\label{sec:GLSH}

\begin{figure}[t]
    \centering
    \includegraphics[width=0.99\linewidth]{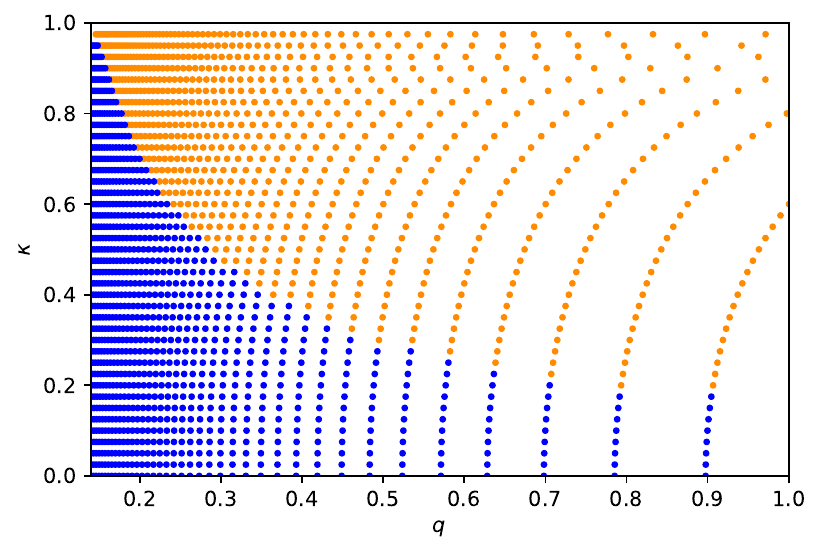}
    \caption{A subregion of $q$-$\kappa$ parameter space depicting the dynamical stability of some generalized longitudinal spin helices in spin-$S$ XYZ chains about the unperturbed point $\delta J_x = 0$.
    Orange dots denote helix parameters exhibiting asymmetric decay (a non-zero Lyapunov exponent of the dynamical matrix $D_k$ to within numerical accuracy for $\delta J_x = -0.01$ but not $\delta J_x = 0.01$; i.e., U-S case) while blue dots denote helix parameters exhibiting symmetric decay (zero Lyapunov exponents of the dynamical matrix $D_k$ to within numerical accuracy for $\delta J_x = \pm 0.01$; i.e., S-S case). 
    Converged results approximating the thermodynamic limit are obtained by numerically diagonalizing $D_k$ in Eq.~\eqref{eq:gshcanonical} for 400 $k$-points to estimate the largest Lyapunov exponent for given $\kappa$ and $q = 4K(\kappa)/\lambda$, where unit cell sizes $\lambda = 7,8,\ldots,80$ are sufficient to populate this plot.
    } 
    \label{fig:phasedia}
\end{figure}

Finally, we consider the generalized longitudinal spin helices. 
In contrast to the cases examined so far, we find that the physics is richer: the presence or absence of asymmetric decay with respect to the perturbation $\delta \hat{H} = \sum_{j=1}^L \delta J_x \hat{S}^x_j \hat{S}^x_{j+1}$ now depends sensitively on the parameters $q,\kappa$ of the underlying longitudinal spin helix (equivalently, the particular unperturbed parent XYZ Hamiltonian).

We numerically compute the spectrum of the dynamical matrices Eqs.~\eqref{eqn:Dk1}, \eqref{eqn:Dk2}, for  small perturbations $\delta J_x = \pm 0.01$ on either side of the unperturbed point $\delta J_x = 0$. We inquire whether the resulting frequency spectrum is purely real (i.e., stable [S]) or contains at least one pair of complex modes signifying instability (i.e., unstable [U]).
In principle, there are four scenarios: S-S, S-U, U-S, U-U, where for example S-U refers to the situation where a perturbation to the {\it left} of the origin (i.e., $\delta J_x < 0$) results in a stable spectrum and a perturbation to the {\it right} of the origin (i.e., $\delta J_x > 0$) results in an unstable spectrum. The other three scenarios are defined analogously.
 In Fig.~\ref{fig:phasedia}, we plot the results of our investigations for a subregion in the $q$-$\kappa$ space defining a generalized longitudinal spin helix. We find, at least within this region, that only the scenarios S-S and U-S occur.
 
\begin{figure}[t]
    \centering
    \includegraphics[width=0.99\linewidth]{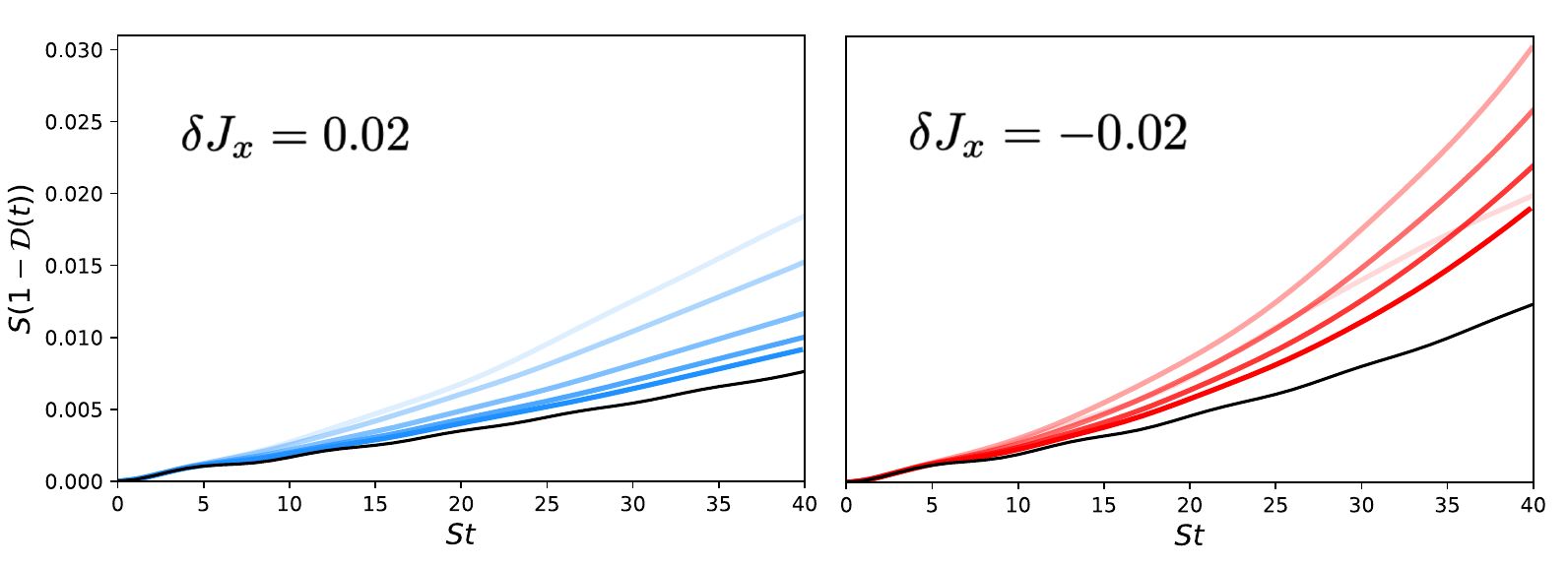}
    \caption{Asymmetric decay of generalized longitudinal spin helices in spin-$S$ XYZ chains. The initial spin texture is given by Eq. \eqref{eq:glsh} with $\kappa = 0.8$ and $q = 4K(\kappa)/7$, corresponding to a wavelength of $\lambda=7$. This Granovskii-Zhedanov scar is an exact product eigenstate when the XYZ couplings $J_x=0.732...$, $J_y=1$ and $J_z = 0.524...$ but decays asymmetrically in perturbations $\delta J_x$ to this value of $J_x$. We simulate 
   the resulting dynamics of the contrast $\mathcal{D}(t)$ for detuning paramters $\delta J_x = 0.02$ (stable regime, left plot) and $\delta J_x = -0.02$ (unstable regime, right plot), with plotting conventions as in Fig.~\ref{fig:transhelixnumerics}.} \label{fig:longhelixnumerics}
\end{figure}

This implies that there are  many GZ scars of longitudinal type (orange region of Fig.~\ref{fig:phasedia}) that exhibit dramatic and asymmetric decay of their contrast, just like the case of all transverse spin helices and generalized transverse spin helices. An illustrative example of the asymmetric decay exhibited by a longitudinal spin helix is compared to microscopic iTEBD simulations in Fig.~\ref{fig:longhelixnumerics}.
On the other hand, there are interestingly also
many  GZ scars of longitudinal type (blue region of Fig.~\ref{fig:phasedia}) that can be {\it exceptionally stable} to perturbations, in particular showing no asymmetry in their decay of contrast to perturbations. 

\section{Discussion}
\label{sec:discussion}
In this work, we have presented a general theory of dynamical instability of quantum many-body scars that are well-defined in the semiclassical limit, and applied it to the class of scars found by Granovskii and Zhedanov, which are exact product eigenstates of arbitrary  XYZ quantum spin-$S$ chains. 
Our theory predicts rich dynamical behavior, notably that GZ scars generically exhibit a decay to perturbations that is non-analytic and asymmetric in the perturbation strength --- depending on the sign of the perturbation, the dynamics can be slow and linear, or fast and exponential, a dramatic difference.
This reveals physics that lies beyond short-time or perturbative expansions like Fermi's Golden Rule. 
At the same time, we also found certain classes of GZ scars (generalized longitudinal spin helices) that have no asymmetry in their decay dynamics and instead show exceptional stability to perturbations.  
These predictions are corroborated by extensive numerics, which also demonstrates that they are valid even far from the semiclassical limit. 
Our findings challenge our understanding of how quantum many-body scars relax, in particular demonstrating that they may not be as unstable as previously thought.   
%
%

As we now explain, our results also hint at a potentially rich web of connections between quantum many-body scars, discrete-time integrable systems and classical nonlinear dynamics. We first note that Granovskii-Zhedanov scars originate from exact, static solutions to the classical, mean-field Landau-Lifshitz equations, which is why they can be constructed for all values of the spin $S$. However, the existence of these exact classical solutions in turn relies on the interpetation of the static Landau-Lifshitz equations as a so-called `discrete-time integrable system' or `integrable map'~\cite{veselov1988integrable,veselov1991integrable}, with the spatial coordinate along the lattice playing the role of a discrete temporal variable. Viewed in this way, the static Landau-Lifshitz equations can be seen as a $0+1$-dimensional discretization of the famous `Neumann problem' of classical rigid-body dynamics. Remarkably, both the Neumann problem and its temporal discretization are integrable in the sense of the Arnol'd-Liouville theorem~\cite{veselov1988integrable}, and this classical integrability is ultimately responsible for the analytical tractability of Granovskii-Zhedanov scars. The question immediately arises of whether this connection between classical discrete-time integrable systems and quantum many-body scars is more widespread: can any of the plethora of known quantum many-body scars be understood in this way? Conversely, are there as yet undiscovered examples of discrete-time integrable systems that give rise to new, infinite families of quantum scars?

A second striking feature of our analysis is the existence of a controlled semiclassical limit as the spin $S \to \infty$, which allows us to establish a particularly clean connection between the Floquet-Lyapunov spectrum of a given stationary classical solution and the stability of the corresponding quantum many-body scar under quantum time-evolution (Sec.~\ref{sec:gentheory} and Appendix \ref{app:quantclass}), complementing recent related results~\cite{Ho_2019,turner2021correspondence,Cotler_2023,hummel2023genuine,lerose2023theory,ermakov2024periodic,evrard2024quantum,pizzi2024quantumscarsmanybodysystems,muller2024semiclassicaloriginsuppressedquantum,omiya2024quantummanybodyscarsremnants}. On the classical side, the stationary spin textures associated with Granovskii-Zhedanov scars can be viewed as nonlinear waves admitting a discrete translation symmetry~\cite{lakshmanan2011fascinating}, and from this perspective their linear instabilities are a special case of the well-known `modulational instabilities' of such classical nonlinear waves~\cite{benjamin1967disintegration,kivshar1993peierls,lakshmanan2008dynamic}. 

This points towards a generic mechanism for how quantum many-body scars, in Heller's original sense of quantum scarring~\cite{heller1984bound}, can arise: certain chaotic, translation-invariant classical lattice models admit nonlinear wave solutions with definite crystal momentum~\cite{kivshar1993peierls,lakshmanan2011fascinating} and whenever these solutions are modulationally unstable, they define unstable classical periodic orbits. According to Heller's theory~\cite{heller1984bound}, the latter are the classical precursors of quantum scars, provided their Lyapunov time is long compared to their oscillation period. If such orbits exist and if the classical lattice models in question arise as a semiclassical limit of a sequence of quantum lattice models, as in this paper, we expect this sequence of quantum lattice models to host quantum many-body scars. Moreover, asymmetric decay of these scars as a function of the perturbation strength follows from a corresponding asymmetric modulational instability of the underlying classical waveform. 

An important question for future work is understanding the scope of these connections between classical nonlinear waves and quantum many-body scars. Beyond the examples studied in this paper, promising candidates for such physics include spin chains constructed from $SU(N)$ operators and Bose-Hubbard models in the semiclassical regime of large occupancy. Although some forty years have elapsed since Granovskii and Zhedanov's striking discovery, the time seems ripe for a fuller exploration of its consequences.

{\it Acknowledgments}.---D.~B.~would  like to thank Wai-Keong (Dariel) Mok and Daniel K.~Mark for interesting discussions. W.~W.~H.~thanks Joaquin Rodriguez-Nieva for useful past discussions. 
D.~B.~acknowledges the use of the computational resources at the National Supercomputing Centre (NSCC), Singapore. 
W.~W.~H.~is supported by the National Research Foundation (NRF), Singapore, through the NRF Felllowship NRF-NRFF15-2023-0008, and through the National Quantum Office, hosted in A*STAR, under its Centre for Quantum Technologies Funding Initiative (S24Q2d0009). 

\bibliographystyle{apsrev4-PRX}
\bibliography{bibl}

\onecolumngrid
\appendix

\section{Necessary and sufficient conditions for a mean-field state to be an exact eigenstate of the XYZ Hamiltonian, and
proof that GZ states are scars
} 
\label{appendix:proof}
Here we follow Granovskii and Zhedanov~\cite{granovskii1985periodic} to outline the sufficient and necessary conditions for a mean-field state, composed as a product of Bloch coherent states
\begin{align}
|\vec{\bo}\rangle = \otimes_{j=1}^L | \mathbf{\Omega}_j\rangle,
\end{align}
to be an eigenvector of a generic nearest-neighbor  quantum spin-$S$ chain 
\begin{equation}
\hat{H} = \sum_{j=1}^L \hat{\mathbf{S}}_j \cdot J \hat{\mathbf{S}}_{j+1}
\label{eqn:H_A1}
\end{equation}
with eigenvalue $E = \sum_{j=1}^L S^2 J^{zz}_{R,j}$. In the process, we prove our claim that GZ states $|\kappa,q,\gamma,\phi\rangle$, Eq.~\eqref{eqn:scar_states}, are exact eigenstates of the XYZ quantum Hamiltonian, Eq.~\eqref{eq:XYZintro}. Above, $J$ is an arbitrary real, symmetric three-by-three exchange matrix. For simplicity, we ignore boundary conditions for now.  
As mentioned in the paper, the Hamiltonian Eq.~\eqref{eqn:H_A1} can be always written as an XYZ Hamiltonian $H_\text{XYZ}$ by rotating the spin operators to align with the principal axes of $J$.

Similar to Sec.~\ref{sec:gentheory}, we go into a new rotated frame of reference --- here, {\it static} ---- such that the mean-field state is the fully $z$-polarized ferromagnetic state $\ket{\Uparrow} = \otimes_{j=1}^L |\hat{\mathbf{z}}\rangle$. Specifically, we define classical rotation matrices $R_j$ with the property that $\bo_j= R_j \hat{\mathbf{z}}$. From this we can define their spin-$S$ representations $\hat{\mathcal{R}}_j$ (quantum operators) such that $\hat{\mathcal{R}}_j(\mathbf{a}_j \cdot \hat{\mathbf{S}}_j) \hat{\mathcal{R}}^\dagger_j = (R_j\mathbf{a}_j) \cdot \hat{\mathbf{S}}_j$ for all $\mathbf{a}_j\in\mathbb{R}^3$. This yields a quantum rotation operator $\hat{\mathcal{R}} = \prod_{j=1}^L \hat{\mathcal{R}}_j$  which has the property $
|\vec{\bo}\rangle = \hat{\mathcal{R}}(t)\ket{\Uparrow}$. 

In the rotated frame, the Hamiltonian is
\begin{align}
\hat{H}_\text{R}=\hat{\mathcal{R}}^\dagger\hat{H}\hat{\mathcal{R}},
\end{align}
which has induced exchange matrix (now no longer necessarily spatially homogeneous)
\begin{equation}
J_{\mathrm{R},j} := R^T_jJ R_{j+1} = \begin{pmatrix}
J_{\mathrm{R},j}^{xx} & J_{\mathrm{R},j}^{xy} & J_{\mathrm{R},j}^{xz} \\
J_{\mathrm{R},j}^{yx} & J_{\mathrm{R},j}^{yy} & J_{\mathrm{R},j}^{yz} \\
J_{\mathrm{R},j}^{zx} & J_{\mathrm{R},j}^{zy} & J_{\mathrm{R},j}^{zz} 
\end{pmatrix}.
\end{equation}
The eigenstate condition $\hat{H}|\vec{\bo}\rangle = E |\vec{\bo}\rangle$ then becomes
\begin{equation}
\hat{H}_{\mathrm{R}} \ket{\Uparrow} = E \ket{\Uparrow}.
\end{equation}
Focusing on the Hilbert space of sites $j$ and $j+1$ and equating terms with matching spin quantum numbers as in Ref.~\cite{Jepsen_2022}, we find that necessary and sufficient conditions for $\vec{\ket{\bo}}$ to be an eigenvector of $\hat{H}$ with eigenvalue $E = \sum_{j=1}^L S^2 J^{zz}_{R,j}$ are 
\begin{align}
\label{eq:eig1}
J^{xx}_{\mathrm{R},j}-J^{yy}_{\mathrm{R},j}+ i(J^{xy}_{\mathrm{R},j}+J^{yx}_{\mathrm{R},j}) = 0,\\
\label{eq:eig2}
(J^{xz}_{\mathrm{R},j} + J_{\mathrm{R},j-1}^{zx}) + i(J^{yz}_{\mathrm{R},j} + J_{\mathrm{R},j-1}^{zy}) = 0,
\end{align}
 which are precisely the conditions obtained by Granovskii and Zhedanov~\cite{GZEarlier,granovskii1985periodic}. Note that Eq.~\eqref{eq:eig2} holds automatically for static mean-field solutions $\dot{\bo}_{j} \equiv 0$ of the Landau-Liftshitz equations  by the constraint Eq.~\eqref{eq:statcond}, while the remaining non-trivial conditions  Eq.~\eqref{eq:eig1} that need to be satisfied for $|\vec\bo\rangle$ to be a quantum eigenstate are
\begin{align}
\label{eq:prod1}
J^{xx}_{\mathrm{R},j} &= J^{yy}_{\mathrm{R},j}, \\ 
\label{eq:prod2}
J^{xy}_{\mathrm{R},j} &= -J^{yx}_{\mathrm{R},j}.
\end{align}

Through direct computation, it can be checked that the mean field GZ states  $|\kappa,q,\gamma,\phi\rangle$, Eq.~\eqref{eqn:scar_states}, satisfy Eqs.~\eqref{eq:eig1} and \eqref{eq:eig2} when $\hat{H}$ is the  XYZ quantum Hamiltonian Eq.~\eqref{eq:XYZintro}, with $J_x,J_z$ related to $\kappa, q$   by Eq.~\eqref{eqn:kappa_squared}, \eqref{eqn:q_eqn}.
This thus establishes that the GZ states $|\kappa,q,\gamma,\phi\rangle$ are indeed quantum many-body scars, as claimed. 

Furthermore, a comparison of the Granovskii-Zhedanov conditions Eqs.~\eqref{eq:eig1}, \eqref{eq:eig2}  with the form of the spin-wave Hamiltonian Eq.~\eqref{eq:HSW}  of the main text reveals a simple physical interpretation: they are equivalent to (i) a {\it static} solution of the classical mean-field Landau-Liftshitz equations of motion, and
(ii)
the {\it absence} of pair creation/annihilation operators in the spin-wave theory Hamiltonian Eq.~\eqref{eq:HSW}, since the former holds iff $\zeta_j=0$ in $\hat{H}_{\mathrm{SW}}$.
In other words, at least as far as the class of Hamiltonians Eq.~\eqref{eqn:H_A1} goes, the existence of a mean-field product state as an exact eigenstate is equivalent to that particular state having no mean-field dynamics and having no quantum fluctuations, at the linearized level. It is perhaps remarkable that these conditions are enough: after all, they are only the first two conditions in an infinite hierarchy of conditions organized in powers of $1/S$ (since in principle, there are additional corrections to quantum dynamics above the spin-wave Hamiltonian).

\section{The spectrum of the  spin-wave Hamiltonian governing quantum fluctuations is equivalent to the spectrum of the classical Hamiltonian governing modulational instability }
\label{app:quantclass}
In this Appendix we explicitly derive the connection between the spectrum of the quantum spin-wave Hamiltonian governing quantum fluctuations on top of the semiclassical trajectory, and the spectrum  of the classical 
Hamiltonian governing modulational instability of the semiclassical trajectory claimed in the main text. The starting point of our analysis  is the general nearest-neighbour spin-$S$ Hamiltonian $\hat{H} = \sum_{j=1}^L \hat{\mathbf{S}}_j \cdot J \hat{\mathbf{S}}_{j+1}$.  

We first derive the classical mean-field equation of motion Eq.~\eqref{eq:MFEoM}. This is most easily done~\cite{balakrishnan1985nonlinear} by projecting the Heisenberg equations of motion for $\hat{\mathbf{S}}$ into the space of products of Bloch coherent states $|\vec{\bo}(t)\rangle = \otimes_{i=1}^L | \bo_i(t)\rangle$, whose defining property for a single spin is~\cite{lieb1973classical}  $\langle{\bo} | \hat{S}^{\alpha} |{\bo} \rangle = S \Omega^{\alpha}$
with ${\bf \Omega} \in \mathbb{R}^3$ and $\| \bo \| = 1$. This yields
\begin{equation}
\partial_t \langle \vec{\bo}(t)| \hat{S}^\alpha_j |\vec{\bo}(t) \rangle  = \langle \vec{\bo}(t) | i[\hat{H},\hat{S}^\alpha_j] | \vec{\bo}(t) \rangle,
\end{equation}
which implies the classical equations of motion
\begin{equation}
\dot{\bo}_i = \frac{\partial H^{\mathrm{cl}}}{\partial \bo_i} \times \bo_i
\end{equation}
for the discrete Landau-Lifshitz Hamiltonian
\begin{equation}
H^{\mathrm{cl}} = S \sum_{j=1}^L \bo_j \cdot J \bo_{j+1},
\end{equation}
recovering Eq.~\eqref{eq:MFEoM}.

To proceed further, we transform to rotating-frame variables $\mathbf{v}_j(t)$ such that $\bo_j(t) = R_j(t) \mathbf{v}_j(t)$, as in Section \ref{sec:gentheory}, which yields the equations of motion
\begin{equation}
\dot{\mathbf{v}}_j = \frac{\partial H^{\mathrm{cl}}_{\mathrm{R}}(t)}{\partial \mathbf{v}_j} \times \mathbf{v}_j
\end{equation}
with time-dependent Hamiltonian
\begin{equation}
\label{eq:classrotH}
H^{\mathrm{cl}}_{\mathrm{R}}(t) = \sum_{j=1}^L S\mathbf{v}_j J_{\mathrm{R},j}(t) \mathbf{v}_{j+1} + \mathbf{h}_{\mathrm{R},j}(t) \cdot \mathbf{v}_j
\end{equation}
in the rotating frame. 
Now consider perturbations $\bo_j(t) = \bo^{(0)}_j(t) + \delta \bo_j(t)$ about the exact mean-field trajectory $\bo^{(0)}_j(t)$, whose dynamics is generated by the classical equations of motion beginning from a choice of perturbation $\delta \bo_j(0)$.
In the rotating frame,  we can write $\mathbf{v}_j = \hat{\mathbf{z}} + \mathbf{w}_j$ for the corresponding perturbation in the rotating frame, where $R_j(t) \mathbf{w}_j =\delta \bo_j(t)$. We find that the linearized dynamics is given by
\begin{equation}
\dot{\mathbf{w}}_j = S(J_{j-1,\mathrm{R}}^T(t) \mathbf{w}_{j-1} + J_{j,\mathrm{R}}(t) \mathbf{w}_{j+1}) \times \hat{\mathbf{z}} + (S(J_{\mathrm{R},j-1}^T(t) + J_{\mathrm{R},j}(t))\hat{\mathbf{z}} + \mathbf{h}_{\mathrm{R},j}(t)) \times \mathbf{w}_j
\end{equation}
while the normalization constraint on the unit vector $\mathbf{v}_j$ further imposes the constraint $\hat{\mathbf{z}}\cdot \mathbf{w}_j=0$. Note also that by definition of $R_j(t)$, the coupling constants in the rotating frame satisfy the stationarity condition
\begin{equation}
\hat{\mathbf{z}} \times (S(J_{\mathrm{R},j-1}^T(t) + J_{\mathrm{R},j}(t))\hat{\mathbf{z}} + \mathbf{h}_{\mathrm{R},j}(t)) = 0
\end{equation}
which recovers Eq. \eqref{eq:statcond}. Thus
\begin{equation}
\dot{\mathbf{w}}_j = \hat{\mathbf{z}} \times \left(\omega_j(t) \mathbf{w}_j - S(J_{j-1,\mathrm{R}}^T(t) \mathbf{w}_{j-1} + J_{j,\mathrm{R}}(t) \mathbf{w}_{j+1})\right),
\end{equation}
where $\omega_j(t)$ was defined in Eq. \eqref{eq:defomega} of the main text. Writing $\mathbf{w}_j = \begin{pmatrix}
q_j \\ p_j \\ 0
\end{pmatrix}$, we find that
\begin{align}
\label{eq:classx}
\dot{q}_j &= -\omega_j(t) p_j + S\left(J^{xy}_{j-1,\mathrm{R}}(t) q_{j-1} + J^{yx}_{j,\mathrm{R}}(t) q_{j+1} + J^{yy}_{j-1,\mathrm{R}}(t) p_{j-1} + J^{yy}_{j,\mathrm{R}}(t) p_{j+1}\right),\\
\label{eq:classy}
\dot{p}_j &= \omega_j(t) q_j - S\left(J^{xx}_{j-1,\mathrm{R}}(t) q_{j-1} + J^{xx}_{j,\mathrm{R}}(t) q_{j+1} + J^{yx}_{j-1,\mathrm{R}}(t) p_{j-1} + J^{xy}_{j,\mathrm{R}}(t) p_{j+1}\right).
\end{align}
These can be written in time-dependent Hamiltonian form
\begin{equation}
\dot{q}_j = \frac{\partial H^{\mathrm{cl}}_{\mathrm{SW}}(t)}{\partial p_j}, \quad \dot{p}_j = -\frac{\partial H^{\mathrm{cl}}_{\mathrm{SW}}(t)}{\partial q_j}
\end{equation}
where the `classical spin-wave Hamiltonian'
\begin{align}
\nonumber H^{\mathrm{cl}}_{\mathrm{SW}}(t) = \sum_{j=1}^L &S\left(J^{xx}_{\mathrm{R},j}(t)q_j q_{j+1} + J^{yy}_{\mathrm{R},j}(t)p_j p_{j+1} + J_{\mathrm{R},j}^{xy}(t)q_jp_{j+1} + J_{\mathrm{R},j}^{yx}(t) p_j q_{j+1}\right) \\
&-\frac{1}{2}\omega_j(t)(q_j^2+p_j^2).
\end{align}
Finally defining complex variables $\alpha_j = \frac{1}{\sqrt{2}}(q_j + ip_j)$, we have
\begin{equation}
i\dot{\alpha}_j = \frac{\partial H^{\mathrm{cl}}_{\mathrm{SW}}}{\partial \alpha^*_j},
\end{equation}
and
\begin{align}
\label{eq:classSW}
H^{\mathrm{cl}}_{\mathrm{SW}}(t) =  \sum_{j=1}^L \eta_j(t) \alpha_{j+1}^*\alpha_j + \eta_j^*(t) \alpha_j^* \alpha_{j+1} + \zeta_j(t) \alpha^*_j \alpha^*_{j+1} + \zeta_j^*(t) \alpha_j \alpha_{j+1} + V_j(t)\alpha_j^* \alpha_j,
\end{align}
where the coefficients $\{\eta_j(t),\zeta_j(t),V_j(t)\}$ are precisely as in Eqs.~\eqref{eq:SWcoupl1}-\eqref{eq:SWcoupl3}. It follows that whenever the underlying mean-field solution $\bo_j(t)$ is periodic in time, the Floquet-Lyapunov spectrum for the classical periodic orbit $\bo_j(t)$ will coincide with the Floquet spectrum of the spin-wave Hamiltonian Eq.~\eqref{eq:HSW}. In particular, the largest Lyapunov exponent is easily obtained as the largest imaginary frequency resulting from the linear dynamics Eqs.~\eqref{eq:classx} and \eqref{eq:classy}, which is in turn equivalent to the Heisenberg-picture quantum dynamics Eq.~\eqref{eq:linearsys}.

\section{Rotated frame conventions and exchange matrices}
\label{app:rotframe}
It will be convenient to consider distinct conventions for (i) transverse spin helices (ii) generalized transverse spin helices with $\Omega^z_j=0$ (iii) generalized longitudinal helices with $\Omega^x_j=0$. Our choices of rotation matrices are certainly not unique and we maintain the following fixed conventions throughout the text.
\subsection{Transverse spin helices}
To model transverse spin helices Eq. \eqref{eq:tsh}, it is simplest to note that for a mean-field solution $\vec{\bo}(t)$ with spherical coordinates
\begin{equation}
\vec{\bo}(t) = \begin{pmatrix}
    \sin{\theta_j(t)} \cos{\varphi_j(t)} \\ \sin{\theta_j(t)} \sin{\varphi_j(t)} \\ \cos{\theta_j(t)}
\end{pmatrix}
\end{equation}
where $0 < \theta_j(t) < \pi$ for all time, the rotation matrix
\begin{equation}
R_j(t) = \begin{pmatrix}
\cos{\theta_j(t)}\cos{\varphi_j(t)} & - \sin{\varphi_j(t)} & \sin{\theta_j(t)}\cos{\varphi_j(t)} \\
\cos{\theta_j(t)} \sin{\varphi_j(t)} & \cos{\varphi_j(t)} & \sin{\theta_j(t)} \sin{\varphi_j(t)} \\
-\sin{\theta_j(t)} & 0 & \cos{\theta_j(t)}
\end{pmatrix}
\end{equation}
consisting of a rotation by $\theta_j(t)$ about the $y$-axis followed by a rotation of $\varphi_j(t)$ about the $z$-axis satisfies $\bo_j(t) = R_j(t) \mathbf{z}$. Making the substitutions $\theta_j(t) = \theta = \mathrm{const}$, $\varphi_j(t) = qj-\omega t$ to match Eq. \eqref{eq:tsh}, we find that the exchange matrix
\begin{equation}
J_{\mathrm{R},j}(t) = J_{\mathrm{R}} = \begin{pmatrix}
\cos{q} + \sin^2{\theta}\,\delta J_z & -\cos{\theta}\sin{q} & -\cos{\theta} \sin{\theta} \,\delta J_z \\
\cos{\theta}\sin{q} & \cos{q} & \sin{\theta}\sin{q} \\
-\cos{\theta}\sin{\theta}\,\delta J_z & - \sin{\theta} \sin{q} & \cos{q} +  \cos^{2}{\theta}\,\delta J_z
\end{pmatrix}
\end{equation}
and the effective magnetic field
\begin{equation}
\mathbf{h}_{\mathrm{R},j}(t) = \mathbf{h}_{\mathrm{R}}= \begin{pmatrix}  2S \cos{\theta}\sin{\theta}\,\delta J_z  \\ 0 \\ -2S \cos^2{\theta}\,\delta J_z \end{pmatrix}
\end{equation}
are both homogeneous in space and independent of time.
\subsection{Generalized transverse spin helices}
For generalized transverse spin helices Eq. \eqref{eq:gtsh}, the rotation matrix
\begin{equation}
R_j = \begin{pmatrix} 0 & - \sn(qj,\kappa) & \cn(qj,\kappa) \\
0 & \cn(qj,\kappa) & \sn(qj,\kappa) \\
-1 & 0 & 0
\end{pmatrix},
\end{equation}
corresponding to a $\pi/2$ rotation about the $y$-axis followed by a rotation by the Jacobi amplitude~\cite{weisstein2002jacobi} $\mathrm{am}(qj,\kappa)$ about the $z$-axis satisfies the desideratum $\bo_j = R_j \mathbf{z}$. This yields the exchange matrix
\begin{equation}
J_{R,j} = \begin{pmatrix} J_z & 0 & 0 \\ 0 & J_y \cn(q,\kappa) & J_y \sn(q,\kappa)\dn(qj,\kappa) \\ 0 & -J_y \sn(q,\kappa)\dn(q(j+1),\kappa) & \frac{J_y\left(\cn(q,\kappa)\dn(q,\kappa) + \cn(qj,\kappa)\sn(qj,\kappa)\dn(qj,\kappa) \kappa^2 \sn^3(q,\kappa)\right)}{1-\kappa^2 \sn^2(qj,\kappa) \sn^2(q,\kappa)} \end{pmatrix}
\end{equation}
at each site and effective magnetic fields $\mathbf{h}_{\mathrm{R},j}=0$ by time independence. 
\subsection{Generalized longitudinal spin helices}
For generalized longitudinal spin helices Eq. \eqref{eq:glsh}, the rotation matrix 
\begin{equation}
R_j = \begin{pmatrix} 1 & 0 & 0 \\ 0 & \dn(qj,\kappa) & \kappa \sn(qj,\kappa) \\ 0 & -\kappa \sn(qj,\kappa) & \dn(qj,\kappa) \end{pmatrix},
\end{equation}
corresponding to a rotation by $-\sin^{-1}(\kappa \sn(qj,\kappa))$ about the $x$-axis satisfies $\bo_j = R_j \mathbf{z}$. This yields the exchange matrix 
\begin{equation}  
J_{R,j} = \begin{pmatrix} J_x & 0 & 0 \\
0 & J_y \dn(q,\kappa) & \kappa J_y \sn(q,\kappa) \cn(qj,\kappa) \\ 0 & -\kappa J_y \sn(q,\kappa) \cn(q(j+1),\kappa) & \frac{J_y\left(\cn(q,\kappa)\dn(q,\kappa) + \cn(qj,\kappa)\sn(qj,\kappa)\dn(qj,\kappa) \kappa^2 \sn^3(q,\kappa)\right)}{1-\kappa^2 \sn^2(qj,\kappa) \sn^2(q,\kappa)}
\end{pmatrix}
\end{equation}
at each site and effective magnetic fields $\mathbf{h}_{\mathrm{R},j}=0$ by time independence. 

\section{Asymptotics of the contrast for transverse spin helices}
\label{app:asymp}
Eq. \eqref{eq:mainresult} of the main text follows from the asymptotics of the scaling function $f(\tau)$ in Eq. \eqref{eqn:f_scaling}, which reads 
\begin{equation}
\label{eq:scalingasymp}
f(\tau) \sim \begin{cases}
\gamma_1 \tau, & -\frac{\cos{q}}{\sin^2{\theta}} <\delta J_z < 0, \\
\frac{A}{\sqrt{\tau}} e^{\tilde{\gamma}_2 \tau}, & \hspace{1.1cm} \delta J_z>0,
\end{cases},\quad \tau \to \infty,
\end{equation}
where $A$ is a non-universal, $S$-independent constant, $\gamma_1$ and $\gamma_2$ are as in Eqs. \eqref{eq:gamma1} and \eqref{eq:gamma2}, and $\tilde{\gamma}_2 = \gamma_2/S$ is independent of $S$. We now derive Eq. \eqref{eq:scalingasymp}.

\subsection{Stable regime}
\subsubsection{An instructive integral}
Consider integrals of the form
\begin{equation}
\label{eq:toyintegral}
I(t) = \int_{-\pi}^\pi dk \, \frac{a(k)}{b(k)} \sin{(2b(k)t)},
\end{equation}
where $a(k)$ is even in $k$ and $b(k)$ is an odd, differentiable function satisfying the following three properties for $k\in [-\pi,\pi]$:
\begin{enumerate}
    \item $b(k) = 0$ iff $k=0$,
    \item $b'(0) > 0$,
    \item $b(k)$ has finitely many stationary points $k^* \neq 0$ with maximum order~\footnote{We say that $p$ is the order of a stationary point $k^*$ if $b^{(p)}(k^*) \neq 0 $ but $b^{(j)}(k^*) = 0$ for $j = 1,2,\ldots,p-1$} $p < \infty$.
\end{enumerate}
We claim that subject to these assumptions,
\begin{equation}
\label{eq:toyasymp}
I(t) \sim \frac{\pi a(0)}{b'(0)}, \quad t \to \infty.
\end{equation}
In order to show this, let $\epsilon > 0$ such that $b(k)$ is strictly increasing on $[-\epsilon,\epsilon]$. Then we can write
\begin{equation}
I(t) = \int_{-\epsilon}^{\epsilon} dk \, \frac{a(k)}{b(k)} \sin{(2b(k)t)} + 2 \int_{\epsilon}^{\pi} dk \, \frac{a(k)}{b(k)} \sin{(2b(k)t)}.
\end{equation}
Write these integrals as $I_1(t)$ and $I_2(t)$ respectively. By our assumptions on $b(k)$ and the method of stationary phase~\cite{bender2013advanced},
\begin{equation}
\label{eq:I2statphase}
I_2(t)  \sim \mathcal{O}(t^{-1/p}), \quad t \to \infty.
\end{equation}
Next consider $I_1(t)$. By monotonicity, $b$ is invertible on $[-\epsilon,\epsilon]$ and we can change variables to $y = 2 b(k)$, yielding
\begin{equation}
I_1(t) = \int_{-2b(\epsilon)}^{2b(\epsilon)} \frac{dy}{2b'(k(y))} \,  a(k(y)) \frac{\sin{(y t)}}{y/2} = \pi  \int_{-2b(\epsilon)}^{2b(\epsilon)} dy \, \frac{a(k(y))}{b'(k(y))} \frac{\sin{(y t)}}{ \pi y}. 
\end{equation}
However, $\frac{\sin(yt)}{\pi y}$ is a nascent delta function and thus
\begin{equation}
\label{eq:I1statphase}
I_1(t) \sim \pi  \int_{-2b(\epsilon)}^{2b(\epsilon)} dy \, \frac{a(k(y))}{b'(k(y))} \delta(y) = \frac{\pi a(0)}{b'(0)}, \quad t \to \infty.
\end{equation}
Eq. \eqref{eq:toyasymp} follows upon combining Eqs. \eqref{eq:I1statphase} and \eqref{eq:I2statphase}.

\subsubsection{Application to the scaling function}
The scaling function $f(\tau)$ satisfies
\begin{equation}
f'(\tau) = \frac{1}{2\pi} \int_{-\pi}^\pi dk \, \left(\frac{\tilde{A}_k^2}{\tilde{w}_k^2}\right) \sin{(2\tilde{w}_k\tau)}.
\end{equation}
We now make the identifications $a(k) = \tilde{A}_k^2$ and $b(k) = \tilde{w}_k$ in Eq. \eqref{eq:toyintegral}. It is clear that $a(k)$ is even and that $b(k)$ is odd. It remains to check conditions 1-3 above. It is easily verified directly that for the parameter values under consideration, the unique zero of $b(k)$ in $[-\pi,\pi]$ is at $k=0$ and that $b'(0) >0$. Similarly, condition 3 holds with $p=2$ for generic allowed values of $\delta J_z$. Then, noting in this case that $a(0) = \sin^{4}{\theta} (\delta J_z)^2$ and $b'(0) = \sqrt{2\cos{q}} \sin{\theta}\, \sqrt{-\delta J_z}$, it follows by Eq. \eqref{eq:toyasymp} that
\begin{equation}
f'(\tau) \sim \frac{1}{2 \pi} \frac{\pi a(0)}{b'(0)} =  \frac{1}{2\sqrt{2}} \frac{\sin^3{\theta}}{\sqrt{\cos{q}}} |\delta J_z|^{3/2} ,\quad \tau \to \infty.
\end{equation}
Thus $f(\tau) \sim \gamma_1 \tau$ as $\tau \to \infty$, where the asymptotic decay rate
\begin{equation}
\label{eq:exactdecay}
\gamma_{1} = \frac{1}{2\sqrt{2}} \frac{\sin^3{\theta}}{\sqrt{\cos{q}}} |\delta J_z|^{3/2}.
\end{equation}
\subsection{Unstable regime}
\subsubsection{An instructive integral}
Consider the integral
\begin{equation}
J(t) = \int_{-\pi}^{\pi} dk \, a(k)\cos{(2b(k)t)},
\end{equation}
where $a(k)$ is even in $k$, bounded and real-valued, and $b(k)$ is odd in $k$, bounded and complex-valued. We further assume that $b(k)$ has the following properties for $k \in [-\pi,\pi]$, where $k_*>0$:
\begin{enumerate}
    \item $b(k) = i b_1(k)$ for $|k| < k_*$, where $b_1(k) \geq 0$ is real-valued.
    \item $b(k) = b_2(k)$ for $|k| > k_*$, where $b_2(k) \geq 0$ is real-valued.
    \item $b(k)=0$ iff $k=0$ or $k = \pm k_*$.
    \item The unique stationary points of $b_1(k)$ for $|k| < k^*$ are maxima at $\pm k^*$, where $0<k^*<k_*$.
    \item Both $a(k)$ and $b_1(k)$ are smooth
    in a sufficiently small neighbourhood of $\pm k^*$.
\end{enumerate}
We claim that subject to these conditions,
\begin{equation}
J(t) \sim a(k^*) \sqrt{\frac{\pi}{|b_1''(k^*)|t}} e^{2b_1(k^*)t}, \quad t \to \infty.
\end{equation}

To see this, first write
\begin{equation}
J(t) = J_1(t) + J_2(t)
\end{equation}
where
\begin{equation}
J_1(t) = 2\int_0^{k_*} dk \, a(k) \cosh{(2b_1(k)t)}
\end{equation}
and
\begin{equation}
J_2(t) = 2\int_{k_*}^{\pi} dk \, a(k) \cos{(2b(k)t)}.
\end{equation}
By boundedness of $a(k)$, $J_2(t)$ is bounded for all time,
\begin{equation}
|J_2(t)| \leq 2 \pi a^*
\end{equation}
where $a^* = \sup_{k \in [-\pi,\pi]} a(k)$. We can also write
\begin{equation}
J_1(t) = \tilde{J}_1(t) + \int_0^{k_*} dk \, a(k) e^{-2b_1(k)t},
\end{equation}
where  $\tilde{J}_1(t) = \int_{0}^{k_*} dk \, a(k) e^{2b_1(k)t}$ and again by boundedness of $a(k)$,
\begin{equation}
|J_1(t) - \tilde{J}_1(t)| \leq \pi a^*.
\end{equation}
Thus we have shown that
\begin{equation}
J(t) \sim \tilde{J}_1(t) + \mathcal{O}(t^0),\quad t \to \infty,
\end{equation}
and it remains to determine the asymptotics of the Laplace-type integral $\tilde{J}_1(t)$. For this, introduce $\epsilon > 0$ sufficiently small and note that up to exponentially small corrections in time~\cite{bender2013advanced}, this integral is dominated by the maximum of $b_1(k)$, as
\begin{equation}
\tilde{J}_1(t) \sim e^{2 b_1(k^*)t}\int_{k^*-\epsilon}^{k^*+\epsilon} dk \, a(k) \exp{\left(- |b_1''(k^*)|(k-k^*)^2 t + 2\sum_{n=3}^{\infty} \frac{b_1^{(n)}(k^*)}{n!} (k-k^*)^n t\right)}.
\end{equation}
Letting $y= (k-k^*)t^{1/2}$ and Taylor expanding $a(k)$ about $k^*$, we find that
\begin{equation}
\tilde{J}_1(t)  \sim \frac{1}{\sqrt{t}}e^{2 b_1(k^*)t}\int_{-\epsilon \sqrt{t}}^{\epsilon \sqrt{t}} dy \, a(k^*+yt^{-1/2}) e^{- |b_1''(k^*)|y^2} \left(1+\mathcal{O}(t^{-1/2})\right) \sim a(k^*) \sqrt{\frac{\pi}{|b_1''(k^*)|t}} e^{2b_1(k^*)t}
\end{equation}
as $t\to \infty$ and the result follows.
\subsubsection{Application to the spin contrast}
Now consider the unstable regime $\delta J_z > 0$. Without loss of generality, we set the branch
\begin{equation}
\tilde{w}_k = 2 \sqrt{2\cos{q}} \sin{(k/2)} (2(\cos{q} + \sin^2{\theta} \, \delta J_z)\sin^2{(k/2)}-\sin^2{\theta}\,\delta J_z)^{1/2}
\end{equation}
of $\tilde{w}_k$ to be in the upper half of the complex plane.

Then there is a non-vanishing window about $k=0$ in which $\tilde{w}_k$ is non-negative and imaginary, namely
\begin{equation}
\sin^2{(k/2)} < \frac{\sin^2{\theta}\,\delta J_z}{2 (\cos{q} + \sin^2{\theta} \, \delta J_z)},
\end{equation}
and the only zero of $\tilde{w}_k$ inside this interval is at $k=0$. The function $\tilde{w}_k$ moreover has two zeros at the boundary of this interval, given by
\begin{equation}
\sin^2{(\pm k_*/2)} = \frac{\sin^2{\theta}\,\delta J_z}{2 (\cos{q} + \sin^2{\theta} \, \delta J_z)}
\end{equation}
with $k_*>0$. We expect the scaling function $f(\tau)$ to be dominated by the value of $\tilde{w}_k$ with largest positive imaginary part in the interval $(-k_*,k_*)$.

In order to make this precise, it will be convenient to focus on the second time derivative
\begin{equation}
\label{eq:contrastsecondderiv}
f''(\tau) = \frac{1}{\pi} \int_{-\pi}^\pi dk \, \tilde{A}_k^2 \cos{(2\tilde{w}_kt)}
\end{equation}
of the scaling function.  Making the identifications $a(k) = \tilde{A}_k^2$ and $b(k) = \tilde{w}_k$ in the previous section, we deduce that 
\begin{equation}
\label{eq:exactunstdecay}
f(\tau) \sim \frac{A}{\sqrt{\tau}} e^{\tilde{\gamma}_2\tau}, \quad \tau \to \infty,
\end{equation}
where $A$ is a non-universal constant and $\tilde{\gamma}_{2} = 2b_1(k^*) > 0$, with $k^*$ the unique maximum of 
\begin{equation}
b_1(k) = 2 \sqrt{2} \sin{(k/2)} \sqrt{\cos{q}}\sqrt{2(\cos{q} + \sin^2{\theta} \, \delta J_z)\sin^2{(k/2)}-\sin^2{\theta}\,\delta J_z} 
\end{equation}
for $0 < k < k_*$. Thus
\begin{equation}
\label{eq:smalldunstable}
\tilde{\gamma}_{2} \approx 2 (\sin^2{\theta}) \delta J_z, \quad 0 < \delta J_z \ll 1
\end{equation}
is linear in $\delta J_z$ in the perturbatively unstable regime.

Note that by the discussion in Appendix \ref{app:quantclass}, $\gamma_2 = S\tilde{\gamma}_2 = 2\lambda^*$ is twice the largest Lyapunov exponent of the transverse spin helix under the classical mean-field dynamics.

\section{Details of numerical simulations}
\label{app:numerical_cal}
In this work, we performed numerical calculations using the infinite time-evolving block decimation (iTEBD) algorithm,   implemented using the TenPy library\,\cite{TenPy}.
TEBD is subject to two systematic and controllable sources of error.
The first source of error stems from approximating the unitary time-evolution operator using the Trotter–Suzuki decomposition\,\cite{TrotterSuzuki}.
This approximation introduces a Trotter error that accumulates over time, scaling as $O(T\delta\tau^2)$ when employing a third-order Trotter-Suzuki decomposition, where $T$ is the total evolution time and $\delta\tau$ is the discrete time step (or Trotter step).
The second source of error arises from using a finite bond dimension in the matrix product state (MPS) representation of the wavefunction.
To accurately capture the system’s time evolution, the bond dimension (and also the entanglement) must increase linearly with time, at a rate determined by the underlying system Hamiltonian and initial state.
While choosing a tiny Trotter step and a large bond dimension can yield highly accurate results at long times, it comes at the cost of significantly increased computational resources and time. 
Therefore, it is essential to select suitable parameters that balance accuracy and efficiency for the simulations.
We choose the parameter set $(\chi=160,\,\delta\tau=0.00025)$, which yields a satisfactory level of accuracy, as elaborated below.
The simulations in the stable regime exhibit excellent convergence for $\chi=160$ and $\delta\tau=0.00025$, as the contrast for spin $S=2$ and $|\delta J_z|=0.05$ converges within the order of $10^{-3}$ ($10^{-4}$) at time $St=40$ ($St=30$), which we estimate by comparing the results obtained using the parameter sets $(\chi=120,\,\delta\tau=0.001)$ and $(\chi=140,\,\delta\tau=0.0005)$.
On the other hand, in the unstable regime, the convergence of the contrast is a bit poor, as the contrast for $S=2$ and $|\delta J_z|=0.05$ converges within the order of $10^{-2}$ ($10^{-3}$) at time $St=40$ ($St=30$).
The accuracy of the contrasts for spin values $S<2$ and $|\delta J_z|<0.05$ is comparable to or better than that for $S=2$ and $\delta J_z=0.05$, and are therefore not discussed further.
\end{document}